\documentclass[3p, times]{elsarticle}
\usepackage[utf8]{inputenc}
\usepackage[dvipsnames]{xcolor}
\usepackage{amsmath}
\usepackage{amsfonts}
\usepackage{amssymb}
\usepackage{bm}
\usepackage{lineno}
\usepackage{tikz}
\usepackage{setspace}
\usepackage{gensymb}
\usepackage{booktabs}
\usepackage[colorlinks,hidelinks]{hyperref}
\usepackage{multirow}
\usepackage{graphicx}
\usepackage{wrapfig}
\usepackage[capitalize,noabbrev,nameinlink]{cleveref}
\usepackage[caption=false,font=footnotesize]{subfig}

\makeatletter
\def\ps@pprintTitle{%
 \let\@oddhead\@empty
 \let\@evenhead\@empty
 \def\@oddfoot{\centerline{\thepage}}%
 \let\@evenfoot\@oddfoot}
\makeatother

\newcommand{\tablescale}{0.85}

\usepackage{listings}
\lstdefinelanguage{json}{basicstyle=\footnotesize, numbers=none, tabsize=2, showstringspaces=false, columns=fullflexible, breaklines=true, frame=lines, string=[s]{'}{'}, comment=[l]{:\ '},morecomment=[l]{:'}}

\begin{document}

\begin{frontmatter}

\title{
ArchABM: an agent-based simulator of human interaction with the built environment. $CO_2$ and viral load analysis for indoor air quality
}

\cortext[correspondingauthor]{Corresponding author}

\author[VICOMTECH]{Iñigo Martinez}\corref{correspondingauthor}
\ead{imartinez@vicomtech.org}

\author[VICOMTECH]{Jan L. Bruse}
\author[VICOMTECH]{Ane M. Florez-Tapia}
\author[TECNUN,ICDIA]{Elisabeth Viles}
\author[VICOMTECH]{Igor G. Olaizola}

\address[VICOMTECH]{Vicomtech Foundation, Basque Research and Technology Alliance (BRTA), Donostia-San Sebastián 20009, Spain}
\address[TECNUN]{TECNUN School of Engineering, University of Navarra, Donostia-San Sebastián 20018, Spain}
\address[ICDIA]{Institute of Data Science and Artificial Intelligence, University of Navarra, Pamplona 31009, Spain}

\begin{abstract}

Recent evidence suggests that SARS-CoV-2, which is the virus causing a global pandemic in 2020, is predominantly transmitted via airborne aerosols in indoor environments. This calls for novel strategies when assessing and controlling a building's indoor air quality (IAQ). IAQ can generally be controlled by ventilation and/or policies to regulate human-building-interaction. However, in a building, occupants use rooms in different ways, and it may not be obvious which measure or combination of measures leads to a cost- and energy-effective solution ensuring good IAQ across the entire building. Therefore, in this article, we introduce a novel agent-based simulator, ArchABM, designed to assist in creating new or adapt existing buildings by estimating adequate room sizes, ventilation parameters and testing the effect of policies while taking into account IAQ as a result of complex human-building interaction patterns. A recently published aerosol model was adapted to calculate time-dependent carbon dioxide ($CO_2$) and virus quanta concentrations in each room and inhaled $CO_2$ and virus quanta for each occupant over a day as a measure of physiological response. ArchABM is flexible regarding the aerosol model and the building layout due to its modular architecture, which allows implementing further models, any number and size of rooms, agents, and actions reflecting human-building interaction patterns. We present a use case based on a real floor plan and working schedules adopted in our research center. This study demonstrates how advanced simulation tools can contribute to improving IAQ across a building, thereby ensuring a healthy indoor environment.

\end{abstract}

\begin{keyword}
agent-based modeling,
indoor air quality,
building ventilation,
aerosol model,
building design,
simulation
\end{keyword}

\end{frontmatter}
\section{Introduction}\label{introduction}



\subsection{Motivation}

Evidence is growing that the virus SARS-CoV-2 that caused a global pandemic in 2020 can be transmitted via inhalation of virus-containing aerosols \cite{Zhang2020,GA2020,Buonanno2020} and recent studies point towards increased infection risk indoors \cite{Li2021,Blocken2021,Agarwal2021,Azuma2020}. For these reasons, the concept of indoor air quality (IAQ) is currently under scrutiny. IAQ is typically determined by air temperature, humidity, and pollutant concentration in closed environments \cite{Valladares2019,ASHRAE}. Due to SARS-CoV-2's significant impact on population health around the globe, authors are calling for taking into consideration the concentration of airborne pathogens or quanta (viral load, i.e., a physical measure of infectious material being present, \cite{CJ2006}) when evaluating IAQ \cite{L2021,Melikov2020,Barbosa2021}. Further, official bodies and organizations are updating their guidelines, standards, and regulations in this regard \cite{ASHRAE,ECDC-EuropeanCenterforDiseasePreventionandControl2020,WHO-WorldHealthOrganization2021}. It is therefore evident that IAQ will need to play a more important role when designing a new building or adapting an existing building. After all, one room with bad IAQ alone may lead to a significant health risk for building occupants.

IAQ is ultimately a product of human interaction in a closed environment as temperature, humidity, and pollutants will rise with the number of occupants and with time if no measure is taken. Generally, IAQ can be controlled via engineering measures (ventilation, air filtration, air disinfection, larger rooms, etc.) or by applying policies or conventions to regulate human interaction in a room (restrict access, limit residence time, etc.). However, all these measures come at a cost or have external constraints, and it is not always obvious which combination of measures will lead to a cost and energy-effective solution that ensures good IAQ throughout the entire building given its use. Novel tools and innovative approaches are therefore necessary \cite{Agarwal2021} to design future buildings and adapt already existing buildings such that they can provide significantly improved IAQ, thereby ensuring a healthy, comfortable and productive indoor environment.

To address this challenge, we propose ArchABM, a novel agent-based simulator designed to assist architects, engineers, and building managers to estimate adequate room sizes, determine adequate ventilation parameters or test the effect of policies while taking into account IAQ across the entire building as a result of complex human-building interaction patterns. In addition, ArchABM can assist in simulating various scenarios to make more informed decisions. Its agent-based engine allows for simulating complex interaction patterns of agents (i.e., occupants) in various rooms, taking into account daily schedules, policies/conventions, and a random factor of agents deciding when to go where and with whom.

To address the call for an "updated" IAQ taking into account airborne virus transmission, we adapted a recently published aerosol model tuned to SARS-CoV-2 \cite{Peng} and can compute time-dependent carbon dioxide ($CO_2$) concentrations and virus quanta levels \cite{Buonanno2020} produced by occupants in rooms. Further, the amount of $CO_2$ inhaled by an agent throughout the day and the number of quanta inhaled by each agent are calculated to reflect a potential physiological response. Both result from the interaction patterns created by all agents acting and interacting in the defined environment. The $CO_2$ level calculations for a specific room were validated by using a well-documented case study from the literature \cite{Candanedo2016} that reported accurate $CO_2$ concentrations over the course of a day in an office.

ArchABM is flexible with regard to the aerosol model thanks to its modular architecture, which allows for the implementation of further models calculating other metrics of interest such as temperature, humidity, or even transmission metrics for other types of viruses. Furthermore, ArchABM is dynamic in the sense that any number of rooms ("\textit{places}"), agents ("\textit{people}") and actions ("\textit{events}") can be defined in a configuration file in order to reflect various human-building-interaction patterns. Aggregation categories for places ("\textit{building}") and people ("\textit{departments}") exist as well. Relevant examples could include buildings designed for any type of work, hospitals or nursing homes, or buildings from the educational sector, to name a few. Note that one ArchABM simulation simulates how agents (occupants) follow or take part in events (activities) occurring in various places as defined in the configuration file over the course of one day. This means that long-term effects from one day to another are not modelled. However, as there is a random component associated with agents' actions and interactions throughout the day, simulation output parameters such as inhaled $CO_2$ and virus quanta levels for individual agents as well as accumulated $CO_2$ and virus levels for individual places will differ from simulation to simulation, even if the configuration file is not changed.
The simulation engine that powers ArchABM is event-driven, resulting in higher computational performance compared to traditional continuous time-stepped ABMs, as an unnecessary update of all components (agents and their environment) at each time step is avoided -- components are only updated when an event is triggered. This allows running a high number of simulations $S_{run}$ (i.e. simulate a high number of individual days of agents interacting in the given environment) for a given configuration file, enabling in-depth statistical analyses of output parameter distributions when comparing different simulation scenarios.

To demonstrate ArchABM's capabilities, we present a human-building-interaction use case for an office scenario with 14 rooms and 60 agents, based on an real floor plan and close-to-real working schedules adopted in our research center. We thereby investigate the impact of a) building-related measures, b) policy-related measures, and c) a building-policy combined case on IAQ in individual locations and the overall building, as well as their effects on individual people in terms of inhaled $CO_2$ and virus quanta.

ArchABM is ready for use as an open source Python library, is available to the public on the official Python Package Index (PyPI) repository\footnote{\url{https://pypi.org/project/archABM/}} \cite{archABM_pypi} and comes with a full documentation\footnote{\url{https://vicomtech.github.io/ArchABM/}} \cite{archABM_github}. All data generated in this study is openly accessible via Mendeley Data\footnote{\url{https://data.mendeley.com/datasets/cx3byrjx7b/1}} \cite{martinez_building_2021}.

\subsection{Related work}
Agent-based models (ABMs) can simulate actions and interactions of autonomous agents within a pre-defined computational environment to calculate outcomes describing how the overall system behaves \cite{Giabbanelli2017}. Since the onset of the pandemic induced by SARS-CoV-2, many studies have been published employing ABMs to model virus transmission dynamics, thereby mainly focused on simulating the effects of policy changes or strategies such as the implementation of social distancing measures or travel restrictions. Most of these studies, however, focus on larger, macro-scale scenarios, in which agents act and interact within an environment representing an entire country, a region, or a city \cite{Chang2020,Kirpich2021,Inoue2020}.

Regarding indoor environments, studies involving ABMs have predominantly analyzed university buildings and campuses, supermarkets, or public spaces such as museums, yet mainly apply exposure-time and contact-distance-based or traditional compartmental SEIR (Susceptible, Exposed, Infected, Recovered) \cite{CJ2006,Silva2020} virus transmission models to calculate the number of infected people after a given time \cite{DOrazio2020,Simeone,Ying2021,Ronchi2020,Vezeteu2020}. IAQ and, specifically, possible airborne transmission via aerosols is often not taken into account. Some other studies apply (pedestrian) movement models to simulate indoor movement patterns without taking into account virus-related parameters \cite{YeungCho2021,Xu2020}. When simulating indoor environments, only a few recent studies explicitly address airborne transmission via aerosols in their ABM \cite{Antczak2021,Farthing2021,Altamimi2021,Zafarnejad2021}. While these studies use sophisticated aerosol models, most of them focus on modeling one single room as the simulation environment (simplified supermarket, choir practice, restaurant, classroom) without taking into account the complex human-building-interaction patterns that may emerge when agents move around a building to different rooms for carrying out diverse activities while being in contact with other agents. Furthermore, most of these studies investigate the impact of policy-based measures (such as wearing a mask) on infection risk and do not analyze the impact of building-related measures on IAQ, such as natural or mechanical ventilation \cite{ASHRAE}, or varying ventilation rates or room sizes.

In the building domain, ABMs have been employed to model building occupants' energy use \cite{Azar2012,Papadopoulos2016} and IAQ in terms of $CO_2$ and other parameters \cite{Jia2019,Uddin2021,Moglia2017}. However, again these studies are typically constrained to simulating one single room which agents can enter and leave and do not consider possible airborne virus transmission.
ArchABM tries to address these issues by providing an easy-to-use and easy-to-install, fast, and flexible agent-based simulator that can simulate complex human-building-interaction patterns while calculating relevant IAQ parameters that take potential airborne virus transmission into account.

\section{Methods}\label{methodology}

\subsection{Agent-based simulator}\label{agent_simulator}

In this article, a novel event-based multi-agent simulation framework is proposed to simulate complex human interaction patterns with and within the built environment and to calculate IAQ metrics and physiological responses. This section summarizes the key features of the proposed framework and formalizes its main components: \textit{events, places, people} and \textit{aerosol model}.

Agent-based simulators can be implemented in two ways: a) continuous simulation and b) event-based simulation. Continuous simulations have a fixed time-step, and the system state is updated in every step. For these simulations, it is critical to select an appropriate period parameter, which indicates how much time elapses between state updates. Furthermore, these simulations can be highly inefficient, as there may not be any changes from one step to the next. Conversely, in discrete event-based simulations, the system is only updated when a new event occurs. The simulator processes new events in sequential order as they are fired or triggered by the simulated entities or agents. An event-based approach is followed for this work.

\paragraph{Engine} \label{sim_engine}
The simulator was implemented using Python 3.7.9 \cite{python}. An object-oriented paradigm was adopted, where each agent is a class instance. The engine was developed on top of the SimPy 4.0.1  \cite{matloff2008introduction} library, a process-based discrete-event simulation framework. Under this paradigm, \textit{processes} are used to model the behavior of active components, such as users. Processes live in an \textit{environment} and interact with the environment and with each other via \textit{events}. The most important event type for our application is the \textit{timeout}, which allows a process to sleep for the given time, determining the duration of the activity. Events of this type are triggered after a certain amount of simulated time has passed. 

\begin{wrapfigure}{r}{0.35\textwidth}
    \centering
     \includegraphics[width=0.28\textwidth]{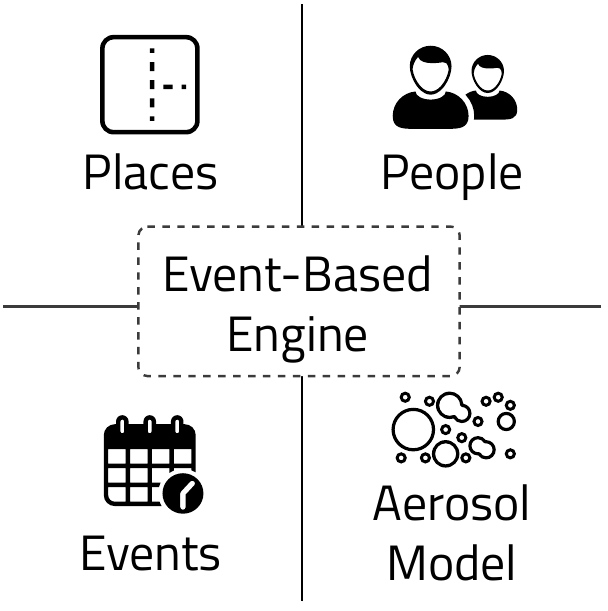}
    \caption{ArchABM main components to be configured: event-based engine, places, people, events and aerosol model}
    \label{fig:architecture}
\end{wrapfigure}

\paragraph{Components} \label{sim_components}
The simulator's core is composed of a discrete event-based engine that manages every activity carried out by the agents during their life-cycle using a priority event queue, ordered by time. The main building blocks of the simulator are depicted in \cref{fig:architecture}. The event-based engine is at the core of the simulator and is fed by events produced by the agents.
During the simulation execution, events are handled sequentially, in chronological order. Whenever any agent does an action or takes a decision, it generates and inserts new events into the priority queue. As actions and activities occur, each event is registered on the simulation history to be further exploited for visualization and data analysis purposes. 



The workflow of the simulator is described as follows: first, Simpy's environment is created, and the provided configuration data is used to generate \textit{events}, \textit{places}, and \textit{people}, as well as to initialize the \textit{aerosol model}. People are introduced into the environment at the start of the day, and their goal is to complete events until the end of the day arrives.

An \textbf{event} is an activity that takes place at a specific physical location for a finite time. Event models (for example: work, meeting, coffee, lunch, etc.) are restricted to a \textit{schedule}, a set \textit{duration}, and a number of \textit{repetitions}. The schedule specifies the times when an activity is permitted to take place. Lower and upper bounds apply to both the duration $\tau$ and the number of repetitions. Concerning the aerosol model, the \textit{mask efficiency} is also indicated for each activity. Activities invoked by an individual but involving many people, such as meetings, can also be defined. These are called \textit{collective} events.

\begin{figure}[!htb]
    \centering
    \resizebox{0.8\linewidth}{!}{
    \begin{tikzpicture}
    \pgfmathsetmacro{\N}{10};
    \pgfmathsetmacro{\M}{6};
    \pgfmathsetmacro{\NN}{\N-1};
    \pgfmathsetmacro{\MM}{\M-1};
    \pgfmathsetmacro{\repmin}{2.25};
    \pgfmathsetmacro{\repmax}{8.5};
    \pgfmathsetmacro{\a}{2};
    
    \coordinate (A) at (0,\MM);
    \coordinate (B) at (\NN,0);
    \coordinate (C) at (\repmin, \a);
    \coordinate (D) at (\repmax, 0);
    \coordinate (E) at (\repmin, 0);
    \coordinate (F) at (0, \a);
    \draw[stepx=1,thin, black!20] (0,0) grid (\N,\M);
    \draw[->, very thick] (0,0) to (\N,0) node[anchor=south west] {\LARGE Event};
    \draw[->, very thick] (0,0) to (\N,0) node[anchor=north west] {\LARGE count $e$};
    \draw[->, very thick] (0,0) to (0,\M)  node[above] {\LARGE Priority};
    \draw (0.1,0) -- (-0.1, 0) node[anchor=east] {\LARGE 0};
    \draw (0, 0.1) -- (0, -0.1);
    \draw (\repmin,0.1) -- (\repmin,-0.1) node[anchor=north, align=center] {\LARGE $r$};
    \draw (\repmax,0.1) -- (\repmax,-0.1) node[anchor=north, align=center] {\LARGE $R$};
    
    \draw[ultra thick] (0.1, \MM) -- (-0.1, \MM) node[left] {\LARGE 1};
    \draw[very thick, black!50, dashed] (C) -- (F) node[left, black] {\LARGE $\alpha$};
    \draw[very thick, black!50, dashed] (C) -- (E);
    \draw[ultra thick, red] (A) -- (C);
    \draw[ultra thick, red] (C) -- (D);
    
    \node[] at (17, 3) {\LARGE Priority$(e) = 
        \left\{\begin{matrix}
        1-(1-\alpha)\cfrac{e}{r}\,,\quad 0 \leq  e < r \\
        \alpha\cfrac{R-e}{R-r}\,,\quad r \leq  e < R \\
        \end{matrix}\right.$};
    \end{tikzpicture}
    }
    \caption{The priority function weighs the importance of each event based on the number of event repetitions $e$.}
    \label{fig:priority}
\end{figure}
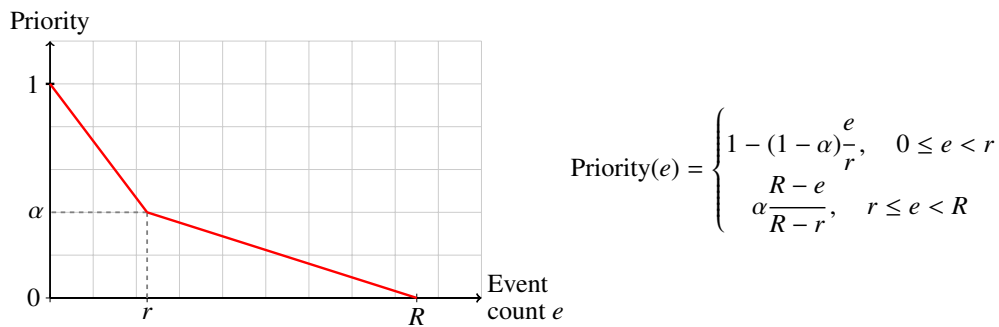

The event generation process selects the next activity based on the priority values of each event model. Priority values are used to weigh the importance of each event model rather than sampling from a uniform discrete distribution. The \textit{priority} value is determined by a piecewise linear function (\cref{fig:priority}), which is parametrized by the a) minimum number of repetitions $r$, b) maximum number of repetitions $R$, and c) the event repetition count $e$. 


A \textbf{place} is an enclosed section of a building designed for specific activities and is defined by the following parameters: building, departments allowed to enter, area and height (or volume), capacity $N$, and natural $\lambda_a$ and mechanical $\lambda_r$ ventilation rates. Note that in the following we understand \textit{natural ventilation} as the introduction of outdoor air into a building driven by naturally produced pressure differentials \cite{ASHRAE}, opposed to \textit{mechanical ventilation}, i.e., recirculation of indoor air by mechanical means such as air conditioning (AC) but without outdoor air supply.

Regarding the \textbf{people} dimension, specific departments or groups need to be defined, each one associated with a building and some people. Finally, the \textbf{aerosol model} estimates the indoor aerosolized virus quanta concentration, based on adjustable parameters such as room size, number of exposed subjects, inhalation volume, and aerosol production from breathing and vocalization, among others. The aerosol model is thoroughly explained in \cref{aerosol_model}.





\begin{figure}[!htb]
    \centering
    \includegraphics[width=0.6\linewidth]{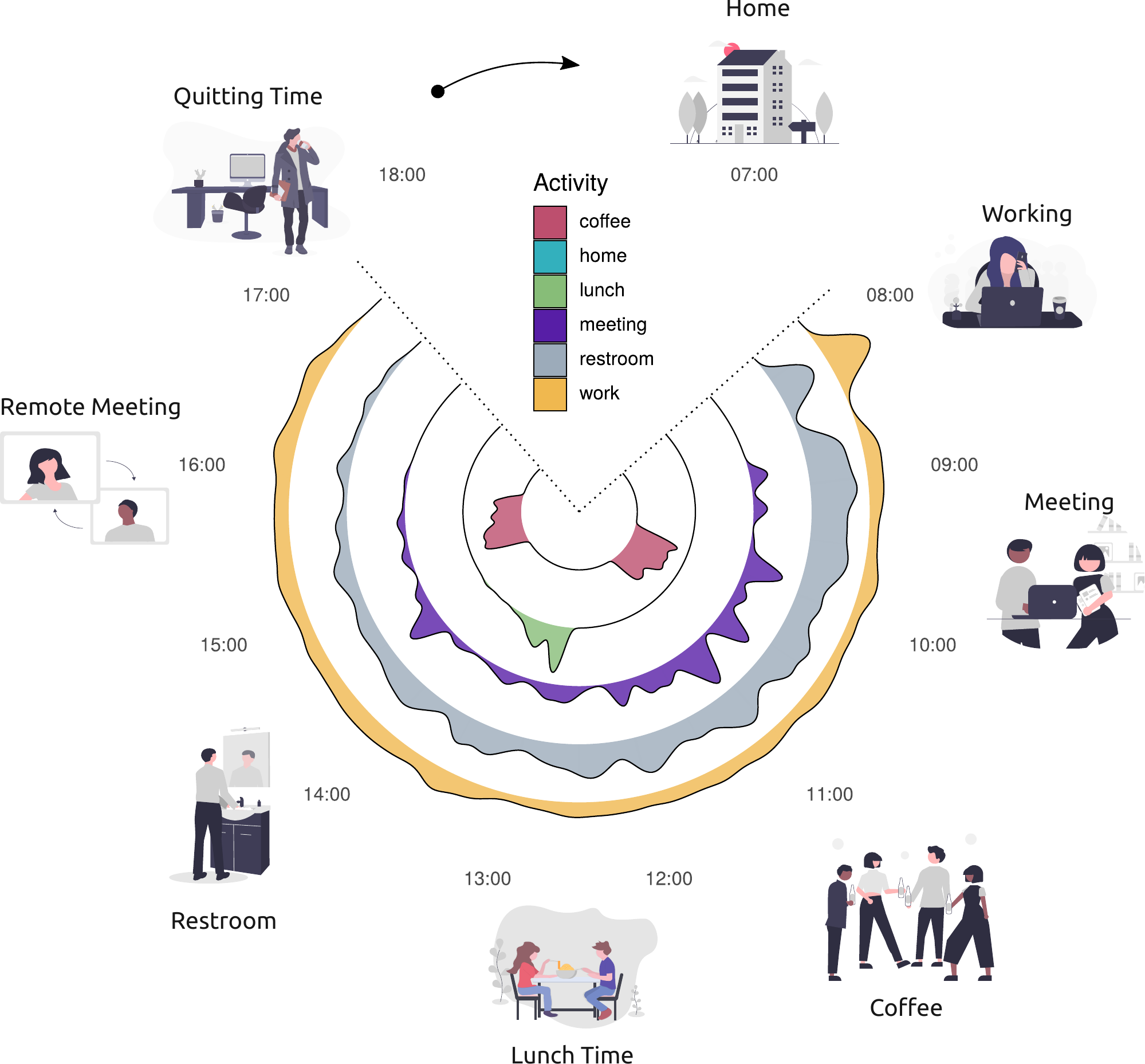}
    \caption{Illustration of a daily schedule example that can be simulated in ArchABM. The radial graph shows the activities' probability density along the day. Colors represent different activities. For instance, lunch is concentrated at noon and coffee breaks during the morning and afternoon, whereas meetings and office work happen uniformly throughout the day.}
    \label{fig:schedule}
\end{figure}

\paragraph{Performance}\label{sim_performance}

\begin{figure}[!htb]
    \centering
    \includegraphics[width=0.5\linewidth]{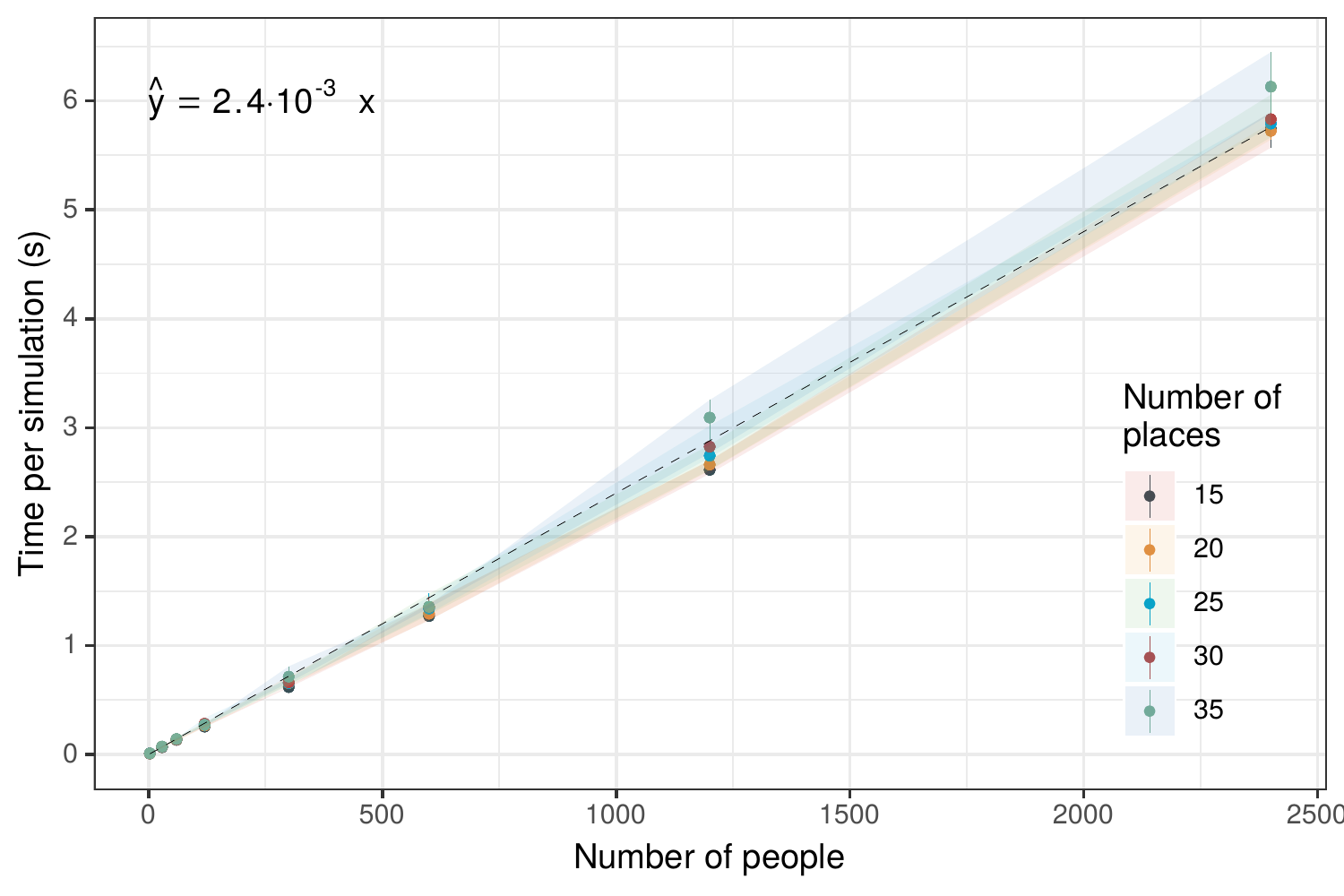}
    \caption{ArchABM computational time performance for different scenarios. The linear regression yields a slope parameter of 2.4 $10^{-3}$ seconds/simulated person.}
    \label{fig:performance}
\end{figure}

In order to analyze ArchABM's computational performance, several simulations were computed with a different number of people and places, as illustrated in \cref{fig:performance}. A grid of values for the number of people  \{6, 30, 60, 120, 300, 600, 1200, 2400\} and the number of places \{15, 20, 25,  30, 35\} was established. The computational time required to compute 24h of simulated time is measured. In order to yield stable results, the simulations are repeated 20 times. The number of people is indeed the most influential parameter concerning the simulator's performance. Using the number of people as the predictor, the univariate linear regression model applied to the response variable time yields a slope parameter of 2.4 $10^{-3}$ seconds per person. Thus, on average, ArchABM is able to run 24h of simulated time with 1000 people and 20 places in approximately 2.4 seconds. 

\subsection{Aerosol model}\label{aerosol_model}

Several models have been proposed to simulate the airborne transmission of SARS-CoV-2, \cite{Bazant,Lelieveld2020,Peng}. Among these models, the model developed by Peng et al. at the University of Colorado, \cite{Peng}, calculates both the virus quanta concentration and the $CO_2$ mixing ratio present in a specific place. The virus quanta is defined as the dose of airborne droplet nuclei required to cause infection in 63\% of susceptible people \cite{Buonanno2020}, and the $CO_2$ mixing ratio refers to the amount of $CO_2$, measured in parts-per-million (ppm), present in the air. These two metrics provide an overall picture of IAQ, which is why this model was selected for ArchABM.

\subsubsection{Standard model by Peng and Jimenez \cite{Peng}}\label{colorado_model}


Recent studies suggest that indoor $CO_2$ measurements hold promise to be used as a proxy for the mass monitoring of indoor aerosol transmission risk for SARS-CoV-2 and other respiratory viruses, \cite{Peng,DiGilio2021,Pang,Bhagat2020}. The aerosol model presented by Peng and Jimenez\cite{Peng} derives analytical expressions of $CO_2$ based risk proxies, assuming the social distance is maintained. The relative infection risk in a given environment scales with excess $CO_2$ level, and thus, keeping $CO_2$ as low as possible is essential to reducing the likelihood of infection. 

The aerosol model \cite{Peng} considers some parameters to be constant across the entire building, as shown in \cref{tab:aerosol_model_constants}. These constant values, breathing or virus related, are based on the study undertaken by Peng and Jimenez, \cite{Peng}. On the contrary, certain parameters are specified for each place, such as volume, ventilation, or the number of people present, as shown in \cref{tab:aerosol_model_variables}.

\begin{table}[!htb]
  \centering
  \scalebox{\tablescale}{%
  \begin{tabular}{lccc}
    \toprule
    & Notation & Units & Value\\
    \midrule
    \multicolumn{4}{l}{\textbf{Building related}}\\
    Background $CO_2$ concentration & $CO_{2(back)}$ & $ppm$ & 415\\
    Pressure & P & $atm$ & 0.95\\
    Temperature & T & $\degree C$ & 20\\
    \midrule
    \multicolumn{4}{l}{\textbf{Breathing related}}\\ 
    Mean breathing flow rate & $Q_b$ & $m^3/h$ & 0.52\\
    $CO_2$ emission rate (1 person) at 273 K and 1 atm & $CO_{2(rate)}$ & $L/s$ & 0.005\\
    \midrule
    \multicolumn{4}{l}{\textbf{SARS-CoV-2 virus related}}\\ 
    Quanta exhalation rate & $Q$ & $quanta/h$ & 25\\
    Decay rate of virus & $k$ & $h^{-1}$ & 0.62\\
    Deposition to surfaces & $\lambda_{dep}$ & $h^{-1}$ & 0.3\\
    Quanta enhancement due to variants & $Q_e$ & - & 1\\
    \midrule
    \multicolumn{4}{l}{\textbf{Simulation related}}\\ 
    Fraction of people using mask & $m_f$ & $\%$ & 100\\
    \bottomrule
  \end{tabular}%
  }
  \caption{Aerosol model parameters: values assumed to be constant across the entire building}
  \label{tab:aerosol_model_constants}
\end{table}

\begin{table}[!htb]
  \centering
  \scalebox{\tablescale}{%
  \begin{tabular}{lcc}
    \toprule
    & Notation & Units\\
    \midrule
    \multicolumn{3}{l}{\textbf{Building related}}\\
    Volume of place & $V$ & $m^3$ \\
    Outdoor, or natural, air exchange rate & $\lambda_a$ & $h^{-1}$\\ 
    Recirculating, or mechanical, air exchange rate & $\lambda_r$ & $h^{-1}$\\ 
    \midrule
    \multicolumn{3}{l}{\textbf{Simulation related}}\\ 
    Mask efficiency & $m_e$ & $\%$ \\
    Number of people & $N$ & - \\
    Number of infected people & $N_i$ & - \\
    Time of exposure & $\tau$ & $h$ \\
    \bottomrule
  \end{tabular}%
  }
  \caption{Aerosol model parameters: variable building- and simulation-related parameters to be defined by the ArchABM user}
  \label{tab:aerosol_model_variables}
\end{table}

From the infection probability point of view, the model considers enclosed spaces, in which virus-containing aerosols are assumed to be rapidly uniformly mixed compared to the time spent by the occupants in the spaces \cite{Peng}.
It states that the probability $p$ for a single person to be infected is related to the number of quanta $n$ of virus inhaled, and according to the Wells-Riley model of aerosol infection \cite{RILEY1962} can be calculated as $p = 1 - e^{-n}$. 
When $p$ is low, as it should be for a safe environment, the use of the Taylor expansion for an exponential allows approximating $p$ as $p \approx n$. 




The $quanta$ inhaled per person $n$ (in $quanta$ units) considers the average $quanta$ concentration $C_{avg}$ [$quanta/m^3$], the mean breathing flow rate $Q_b$ defined in \cref{tab:aerosol_model_variables}, as well as the exposed time $\tau$, the mask efficiency, $m_e$, and the fraction of people using masks, $m_f$:

\begin{equation}\label{eq:n}
n = C_{avg} \cdot Q_b \cdot \tau \cdot (1-m_e\cdot m_f)
\end{equation}

Note that assuming a constant $m_f$ = 100$\%$ = 1 as defined in \cref{tab:aerosol_model_constants} means that one can control whether masks are being used by setting $m_e$ either to 0 (no masks used) or to a value referring to low (0.3), medium (0.5) or high (0.75) mask efficiency, following \cite{Peng} and \cite{Lelieveld2020}.
The aerosol $quanta$ concentration increases with time from an initial value of zero following a $f(x) = 1-e^x$ function, which is the standard dynamic response of a well-mixed indoor volume to a constant input source, \cite{Miller2021}. The average $quanta$ concentration, $C_{avg}$, is calculated as follows: 

\begin{equation}\label{eq:Q_c}
C_{avg} = \frac{E}{\lambda \cdot V} \cdot \left[1-\left(\frac{1-e^{-\lambda \cdot \tau}}{\lambda \cdot \tau}\right)\right]
\end{equation}

where $E$ is the net emission rate in [quanta/h] units, and $\lambda$ the total first order loss rate in [$h^{-1}$] units: 

\begin{equation}\label{eq:E}
E = Q \cdot (1 - m_e \cdot m_f) \cdot N_i \cdot Q_e
\end{equation}

\begin{equation}\label{eq:lambda}
\lambda = \lambda_a + \lambda_r + \lambda_{dep}+ k
\end{equation}

$\lambda_r$ in \cref{eq:lambda} is the recirculating, or mechanical, ventilation, and has to do with the flow rate of the AC system, $Q_{AC}$, the filter efficiency, $\varepsilon_{filter}$, the removal in ducts, $\varepsilon_{ducts}$, and additional removal measures, $\varepsilon_{extra}$:

\begin{equation}\label{eq:lambda_r}
    \lambda_r = \frac{Q_{AC}}{V} \cdot \min \{\varepsilon_{filter} + \varepsilon_{ducts} + \varepsilon_{extra}, 1\}
\end{equation}

The model also calculates the $CO_2$ mixing levels reached within the specified area, given that there are no other significant $CO_2$ sources or sinks, i.e., indoor excess $CO_2$ production, relative to the outdoor background level, is only due to human exhalation and its loss is ventilation, \cite{Peng}:

\begin{equation}\label{eq:CO2}
CO_{2} \leftarrow CO_{2(emit)} \cdot \frac{3.6 \cdot 10^6}{\lambda_a \cdot V} \cdot \left[1-\frac{(1-e^{-\lambda_a \cdot \tau})}{\lambda_a \cdot \tau}\right] + CO_{2(back)}
\end{equation}

The $CO_{2(emit)}$ represents the total amount of $CO_2$ emitted by all people present, and is calculated by taking into account the $CO_2$ emission rate for one person as defined in \cref{tab:aerosol_model_constants}, at nominal temperature and pressure conditions (273 K, 1 atm). This value should be adjusted to the pressure and temperature of the building being simulated and multiplied by the number of people present:

\begin{equation}\label{eq:CO2_rate_all}
CO_{2(emit)} = CO_{2(rate)} \cdot \frac{N}{P} \cdot \frac{(273.15 + T)}{273.15}
\end{equation}

\subsubsection{Extended model for ArchABM}\label{colorado_model_adapted}


The model developed by Peng and Jimenez \cite{Peng} provides equations for a single event of a given duration. However, ArchABM simulates an entire day through short-term events. As a result, there must be some continuity, i.e., a transition from a static model that assumes an initial clean environment towards a dynamic and continuously adapting model that considers how the previous state affects the next state.

According to \cref{eq:Q_c}, the $quanta$ concentration can only increase during an event.
However, in the scenario presented by ArchABM, people are moving between different locations and when no contagious people are present the $quanta$ concentration should decay (see \cref{eq:Q_c_i}), due to the total removal rate, $\lambda$, which takes into account the ventilation rates, $\lambda_a$ and $\lambda_r$, as well as the decay rate of the virus, $k$, along with its deposition rate to surfaces, $\lambda_{dep}$ (see {\cref{eq:lambda}}). \cref{eq:Q_c} was adapted to account for this fact:

\begin{equation} \label{eq:Q_c_i}
C_{avg} \leftarrow \frac{E}{\lambda \cdot V} \cdot \left[1-\left(\frac{1-e^{-\lambda \cdot \tau}}{\lambda \cdot \tau}\right)\right] + C_{avg} \cdot e^{-\lambda \cdot \tau}
\end{equation}

The $quanta$ inhaled per person $n$ is calculated as in \cref{eq:n}, but considering the updated expression for the $quanta$ $C_{avg}$ in \cref{eq:Q_c_i}.

The $CO_2$ level formulated in \cref{eq:CO2} follows a similar adaptation process.
In the original aerosol model, the $CO_2$ concentration can only increase when people are present and are breathing for a specified time.
However, when simulating a day through short-term events, the model should also consider the scenario of these people leaving the room and the effect of the ventilation.
The $CO_2$ concentration rates should decrease if the ventilation continues and the room is empty.
Therefore, the extended $CO_2$ mixing equation now takes the previous state into account:

\begin{equation} \label{eq:CO2_i}
CO_{2}  \leftarrow CO_{2(emit)} \cdot \frac{3.6 \cdot 10^6}{\lambda_a \cdot V} \cdot \left[1-\frac{(1-e^{-\lambda_a \cdot \tau})}{\lambda_a \cdot \tau}\right] + CO_{2(back)} + \left(CO_{2} - CO_{2(back)}\right) \cdot e^{-\lambda_a \cdot \tau} 
\end{equation}


With these modified equations, the ArchABM simulator can estimate the $CO_2$ mixing ratio level [ppm] at each place as well as the $quanta$ concentration $C_{avg}$ [quanta/$m3$], providing an overall picture of the IAQ distribution per room throughout the day. 

\subsection{Simulator input and outputs}\label{sim_in_out}

In order to run a simulation, information about the event types, people, places, and the aerosol model must be provided to the ArchABM framework:
\begin{itemize}
    \item[--] \textbf{Events} input parameters: name (\textcolor{teal}{\textit{string}}), schedule (\textcolor{teal}{\textit{list of tuples}}), duration range (\textcolor{teal}{\textit{integer, integer}}), number of repetitions (\textcolor{teal}{\textit{integer, integer}}), mask efficiency (\textcolor{teal}{\textit{float}}), and collective (\textcolor{teal}{\textit{boolean}}).
    
    \item[--] \textbf{Places} input parameters: name (\textcolor{teal}{\textit{string}}), activity (\textcolor{teal}{\textit{string}}), building (\textcolor{teal}{\textit{string}}), department (\textcolor{teal}{\textit{list of strings}}), area (\textcolor{teal}{\textit{float}}), height (\textcolor{teal}{\textit{float}}), capacity (\textcolor{teal}{\textit{int}}), natural ventilation (\textcolor{teal}{\textit{float}}), and mechanical ventilation (\textcolor{teal}{\textit{float}}).
    
    \item[--] \textbf{People} input parameters: name (\textcolor{teal}{\textit{string}}), building (\textcolor{teal}{\textit{string}}), and department (\textcolor{teal}{\textit{string}}).
    
    \item[--] \textbf{Aerosol model} input parameters: included on \cref{tab:aerosol_model_constants} and \cref{tab:aerosol_model_variables}
\end{itemize}

Regarding ArchABM's output, whenever a new event occurs, the simulator saves the state of each person and each place in the simulation history data structure. The following attributes are stored: 
\begin{itemize}
    \item[--] \textbf{Places} output metrics: place ID (\textcolor{teal}{\textit{int}}), number of people (\textcolor{teal}{\textit{int}}), number of infected people (\textcolor{teal}{\textit{int}}), $CO_2$ level (\textcolor{teal}{\textit{float}}), and $quanta$ level (\textcolor{teal}{\textit{float}}).

    \item[--] \textbf{Person} output metrics: person ID (\textcolor{teal}{\textit{int}}), simulation time (\textcolor{teal}{\textit{float}}), place ID (\textcolor{teal}{\textit{int}}), event ID (\textcolor{teal}{\textit{int}}), $CO_2$ level at current place (\textcolor{teal}{\textit{float}}), and $quanta$ inhaled during the event (\textcolor{teal}{\textit{float}}).
\end{itemize}

In this way, ArchABM provides full tractability of the places visited by every person, their physiological responses as well as the instantaneous IAQ at each place.

\subsection{Experimental setup}\label{experiments_setup}



\begin{figure}[!htb]
    \centering
    \includegraphics[width=0.7\linewidth]{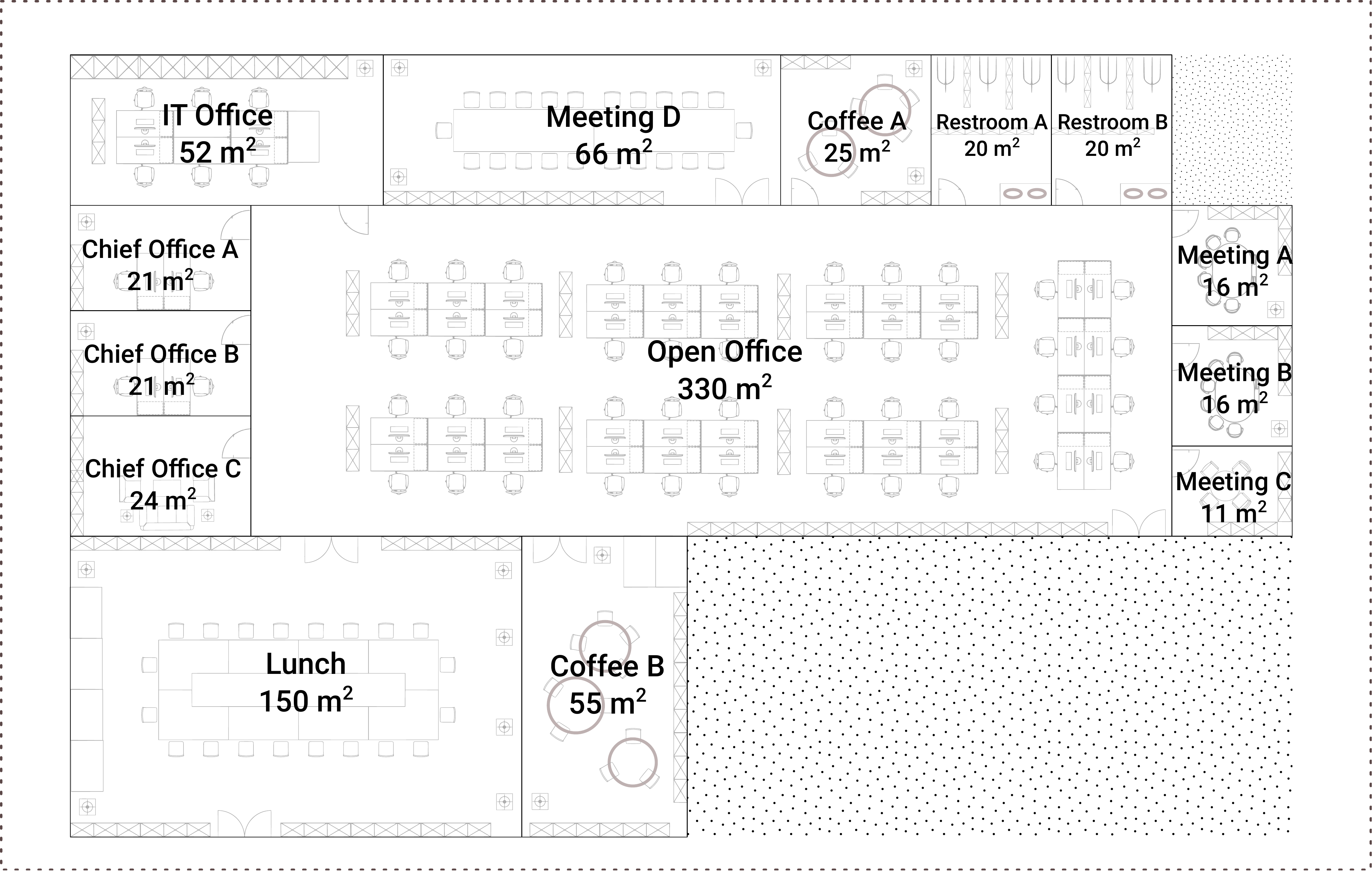}
    \caption{Building floor plan used in the experiments}
    \label{fig:building_plan}
\end{figure}

The simulated configuration is based on the real floor plan of one building of our research center, as shown in \cref{fig:building_plan}. It should be noted that the floor plan is only shown for illustration purposes and is not required to run the simulator. 
As explained in \cref{sim_components}, the simulator only requires information about the types of \textit{events} that can occur, \textit{places}' spatial parameters (area \& capacity), the number of \textit{people} initially present and the \textit{aerosol model} physical parameters. 
The events and places defined for this simulation are summarized in \cref{tab:baseline_events} and \cref{tab:baseline_places}. There are five types of events: \textit{work, meeting, coffee, restroom}, and \textit{lunch}. Meetings and lunch activities are regarded as collective events. Each event model is limited to a certain schedule, duration $\tau$, and a number of repetitions. For each event model, the mask efficiency $m_e$ is also defined.

The floor area of each location was measured, and the volume $V$ was estimated assuming a height of 2.7 m. The initial number of people present $N$, the maximum number of people that can fit in the room, and the ventilation rates (both mechanical $\lambda_r$, i.e., without outdoor air supply, and natural $\lambda_a$, i.e., with outdoor air supply) are also defined. Meeting rooms A, B, C, and restrooms A and B are subject to poor natural ventilation, as they are oriented towards the interior of the building and do not have direct access to a window.

There are 60 people distributed in 7 different departments: D1, D2, and D3 have 16 people each; D4 refers to 7 Information Technology (IT) workers, and D5, D6, D7 hold the head of departments and the Chief Executive Officer (CEO), with 2, 2, and 1 people respectively. Each place is only accessible to people from specific departments. This is determined by the "departments allowed" parameter on {\cref{tab:baseline_places}}. The IT office, for example, is only accessible to members of the IT department D4. Due to accessibility limitations (no unisex bathrooms), one of the restrooms (restroom B) is open to all departments except D7 (CEO). Note that departments D1-D4 can work in the open office for extended periods of time throughout the day, whereas managers (departments D5-D7) can only walk through and cannot work or meet with employees from departments D1-D4 in the open office. The number of infected people $N_i$ is set to 3 in all the proposed scenarios. The aerosol model parameters are described in \cref{colorado_model}.

\begin{table}[!htb]
    \centering
    \scalebox{\tablescale}{%
    \begin{tabular}{lccccc}
    \toprule
    Activity  & Schedule  & \begin{tabular}[c]{@{}c@{}}Duration\\ $\tau$ {[}$h${]}\end{tabular} & \begin{tabular}[c]{@{}c@{}}Repetitions\\ {[}$times${]}\end{tabular} & \begin{tabular}[c]{@{}c@{}}Mask Eff.\\ $m_e$ {[}$\%${]}\end{tabular} & Collective \\
    \midrule
    Work     & 8:00 - 17:00   & 0.5 - 1  & 0 - $\infty$  & 0     & $\times$      \\
    Meeting  & 9:00 - 16:00   & $0.\hat{3}$ - 1.5  & 0 - 5   & 0   & \checkmark       \\
    Coffee   & \begin{tabular}[c]{@{}l@{}}10:00 - 10:30\\ 15:00 - 16:00\end{tabular} & $0.08\hat{3}$ - 0.25 & 0 - 2  & 0   & $\times$ \\
    Restroom & 8:00 - 17:00  & 0.05 - 0.1  & 0 - 4   & 0   & $\times$      \\
    Lunch    & 13:00 - 15:00  & $0.\hat{3}$ - 0.75   & 1 - 1  & 0  & \checkmark      \\
    \bottomrule
    \end{tabular}%
    }
    \caption{Event parameters for the baseline case}
    \label{tab:baseline_events}
\end{table}

\begin{table}[!htb]
    \centering
    \scalebox{\tablescale}{%
    \begin{tabular}{lccccccc}
    \toprule
    & Area {[}$m^2${]} & People $N$  & Capacity {[}people{]} & $\lambda_a$ {[}$h^{-1}${]} & $\lambda_r$ {[}$h^{-1}${]} & \begin{tabular}[c]{@{}c@{}}Departments\\allowed\end{tabular} & Activity  \\
    \midrule
    \multicolumn{7}{l}{\textbf{Offices}} & \multirow{6}{*}{Work}\\
    Open Office    & 330  & 48 & 60   & 1.5  & 0 & D1,D2,D3,D4 &  \\
    IT Office      & 52   & 7  & 10   & 1.5  & 0 & D4 &   \\
    Chief Office A & 21   & 2  & 5    & 1.5  & 0 & D5,D6,D7 & \\
    Chief Office B & 21   & 2  & 5    & 1.5  & 0 & D5,D6,D7 & \\
    Chief Office C & 24   & 1  & 5    & 1.5  & 0 & D5,D6,D7 & \\
    \midrule
    \multicolumn{7}{l}{\textbf{Meeting rooms}} & \multirow{5}{*}{Meeting}\\
    Meeting Room A & 16   & -  & 6    & 0.5  & 0 & All except D4 &  \\
    Meeting Room B & 16   & -  & 6    & 0.5  & 0 & All except D4 &   \\
    Meeting Room C & 11   & -  & 4    & 0.5  & 0 & All except D4 & \\
    Meeting Room D & 66   & -  & 24   & 1.5  & 0 & All & \\
    \midrule
    \multicolumn{7}{l}{\textbf{Coffee rooms}} & \multirow{3}{*}{Coffee}  \\
    Coffee A       & 25   & -  & 10   & 1.5  & 0 & All &  \\
    Coffee B       & 55   & -  & 20   & 1.5  & 0 & All &  \\
    \midrule
    \multicolumn{7}{l}{\textbf{Restrooms}} & \multirow{3}{*}{Restroom}\\
    Restroom A     & 20   & -  & 4    & 0.5  & 0 & All &  \\
    Restroom B     & 20   & -  & 4    & 0.5  & 0 & All except D7 &  \\
    \midrule
    \multicolumn{7}{l}{\textbf{Lunch room}}\\
    Lunch          & 150  & -  & 60   & 1.5  & 0 & All & \multirow{1}{*}{Lunch}  \\
    \bottomrule
    \end{tabular}%
    }
    \caption{Place parameters for the baseline case}
    \label{tab:baseline_places}
\end{table}


\newpage
\subsubsection{Measures}

The following section describes what actions can be implemented to control indoor air quality and reduce $CO_2$ and $quanta$ concentration levels.

\paragraph{Building-related} 

Architects and engineers can increase the space's area and/or height during the design stage and create open spaces or more separated workspaces. The ventilation strategy can also be changed through mechanical systems like air conditioning (AC) or portable high-efficiency particulate air (HEPA) filters, or through natural ventilation, obtaining outdoor air exchange through windows or doors. 

\paragraph{Company policy-related} 

Physical measures include mandatory masks, locking / restricting access to small rooms, or renting additional rooms. Other measures entail reducing the number of people working from the office through remote shifts, reducing the number or duration of meetings, restricting movements between buildings or departments, and, in some cases, prohibiting eating lunch in the office to avoid no-mask scenarios. This latter possibility, however, may be unrealistic.

\subsubsection{Experiments}\label{experiments}


ArchABM can help to quantify the impact of some of these building- and company policy-related measures. Therefore, the proposed experiments are defined in this section. 


\paragraph{Baseline case - no measures}


A baseline case with no measures and reduced ventilation is first studied. The events' parameters are summarized in \cref{tab:baseline_events}, where an \textit{schedule} is set for each event type, along with their minimum and maximum \textit{duration} $\tau$, and the number of \textit{repetitions}. The \textit{mask} is not used anywhere ($m_e = 0$). Meetings and the lunch activity are considered to be \textit{collective} events. 

Places' parameters are summarized in \cref{tab:baseline_events}. The \textit{capacity} refers to the maximum number of people that can be present in that specified space. A low natural ventilation rate is established ($\lambda_a = 1.5$, and $\lambda_a = 0.5$ for poor ventilated rooms) and there is no mechanical ventilation ($\lambda_r = 0$). 


\paragraph{Building-related} 


\begin{enumerate}
    \item Larger building: each room's area (and thus each room's volume) is increased by 20$\%$. This measure needs to take into account the increase in costs, which according to \cite{costmo} would mean an increase of almost $20\%$ in the final construction costs as well. 
    
    \item Separate workspaces: the \textit{open office} is divided into three identical offices, each one with 110 $m^2$, 16 people (48/3), and a capacity of 20 (60/3).
    
    \item Better natural ventilation: windows are opened everywhere except in restrooms for better outdoor air supply. $\lambda_a$ is increased up to 5 $h^{-1}$.
    
    \item Better mechanical ventilation: the flow rate $Q_{AC}$ of the AC system is incremented, assuming a 20\% filter efficiency $\varepsilon_{filter}$, a 10\% of removal in ducts $\varepsilon_{ducts}$ and no additional $\varepsilon_{extra}$ removal measures. According to \cite{costmo}, adding AC to the building would mean an increase of 14\% in the building overall costs. According to \cref{eq:lambda_r}, the values resulting for $\lambda_r$ are summarized in \cref{tab:Incremented active ventilation}:

\begin{table}[!htb]
  \centering
  \scalebox{0.85}{%
  \begin{tabular}{lcc}
    \toprule
    & $Q_{AC}$ [$m^3/h$] & $\lambda_r$ [$h^{-1}$]\\
    \midrule
    \multicolumn{3}{l}{\textbf{Offices}}\\
    Open Office & 1000 & 0.337\\
    IT Office & 300 & 0.641\\
    Chief Office A & 300 & 1.587\\
    Chief Office B & 300 & 1.587\\
    Chief Office C & 300 & $1.3\hat{8}$\\
    \midrule
    \multicolumn{3}{l}{\textbf{Meeting rooms}}\\ 
    Meeting Room A & 300 & $2.08\hat{3}$\\
    Meeting Room B & 300 & $2.08\hat{3}$\\
    Meeting Room C & 300 & $3.\hat{03}$\\
    Meeting Room D & 1000 & 1.68\\
    \midrule
    \multicolumn{3}{l}{\textbf{Coffee}}\\ 
    Coffee A & 300 & $1.\hat{3}$\\
    Coffee B & 1000 & $2.\hat{02}$\\
    \midrule
    \multicolumn{3}{l}{\textbf{Restrooms}}\\ 
    Restroom A & 0 & 0\\
    Restroom B & 0 & 0\\
    \midrule
    \multicolumn{3}{l}{\textbf{Lunch}}\\ 
    Lunch & 1000 & $0.\hat{740}$\\
    \bottomrule
  \end{tabular}%
  }
  \caption{Parameters for a better mechanical ventilation experiment involving AC with air recirculation, but no outdoor air supply}
  \label{tab:Incremented active ventilation}
\end{table}

\end{enumerate}

\paragraph{Policy-related}


\begin{enumerate}
    \item Shifts between workers: this would imply a reduction in the number of people present in each room. For this experiment, the population is reduced by 40\%, resulting in 29 people in the open office, 4 in the IT Office, and 1 in each chief office, summing up to 36 people. This measure also entails a non-quantifiable cost to the company. 
    
    \item Limit duration of events: the duration of meetings is limited to a maximum of 30 minutes, setting $\tau = [0.\hat{3} - 0.5] h$. The duration of coffee breaks would be limited to 5 minutes, meaning $\tau = 0.08\hat{3} h$, and lunch would be of 20 minutes, $\tau = 0.\hat{3} h$.
    
    \item Use of masks: in this case, the mask use is mandatory, meaning that $m_f = 1$ and the mask efficiency, $m_e$, is set to 0.75 in the offices and meeting rooms, to 0.5 in the restrooms, to 0.3 for coffee breaks and leaving it at 0 for lunch breaks, representing the absence of masks while eating. 
\end{enumerate}

\paragraph{Combined case}


Finally, in order to quantify the impact of implementing both building and policy measures, the experiments of \textit{better natural ventilation} and \textit{limit events duration} are combined in a new experiment.

\subsection{Statistical analysis}\label{statistical_analysis}
The results of the described experiments are evaluated with regard to three levels: places, people (i.e., departments), and the entire building. In terms of outcome parameters related to IAQ at the place-level, the maximum $CO_2$ level (concentration in ppm) and the maximum virus $quanta$ level (concentration in ppm) reached during the day per place are calculated. In terms of physiological response outcome at person-level, the time-weighted average inhaled $CO_2$ over the day and the maximum $quanta$ inhaled at the end of the day per person are used. At the building level, volume-weighted average maximum $CO_2$ is reported in terms of IAQ parameters per experiment, where volume refers to the volume of each place. To summarize physiological response parameters on the building level, the maximum $quanta$ level at the end of the day is averaged over all people.

Prior to further statistical analyses of these outcome parameters, we conducted a set of trial simulations to determine an adequate number of simulation runs $S_{run}$. Note that one simulation, i.e. one simulation run refers to simulating agents' actions and interactions in the given environment as defined in the input configuration file over the course of one day.
Certain ABMs are prone to be statistically underpowered as they may require much computational time and effort to complete one single simulation \cite{Seri2017}. This promotes Type-I errors of not detecting an actual effect. ArchABM is computationally efficient and takes less than a second on a standard laptop to complete one simulation with the given baseline configuration, which could encourage running many simulations. This, however, may lead to overpowered analyses, promoting Type-II errors of detecting a non-existent effect.
For this, we ran each of the nine experimental configurations from $S_{run}$ = 10 up to $S_{run}$ = 1000 simulations in unitary steps and repeated each setting 100 times (e.g., we computed a set of 100 simulations for each $S_{run}$). For every single simulation, we computed further the \textit{coefficient of variation} ($CV$) \cite{Lee2015}, defined as the ratio of the standard deviation of a sample to its mean for each of the four critical outcome parameters (maximum $CO_2$ and $quanta$ level per place and mean inhaled $CO_2$ and inhaled $quanta$ per department). Low consecutive $CV$s imply the stability of the results. Therefore, $CV$s were plotted for all simulation sets over $S_{run}$ for all outcome parameters, and convergence of $CV$ towards a stable range was visually assessed. The resulting plots are detailed in the \ref{appendix_CV}. Following this analysis, the adequate $S_{run}$ was 500 for all experiments.

\begin{figure}[!htb]
    \centering
    \includegraphics[width=0.7\linewidth]{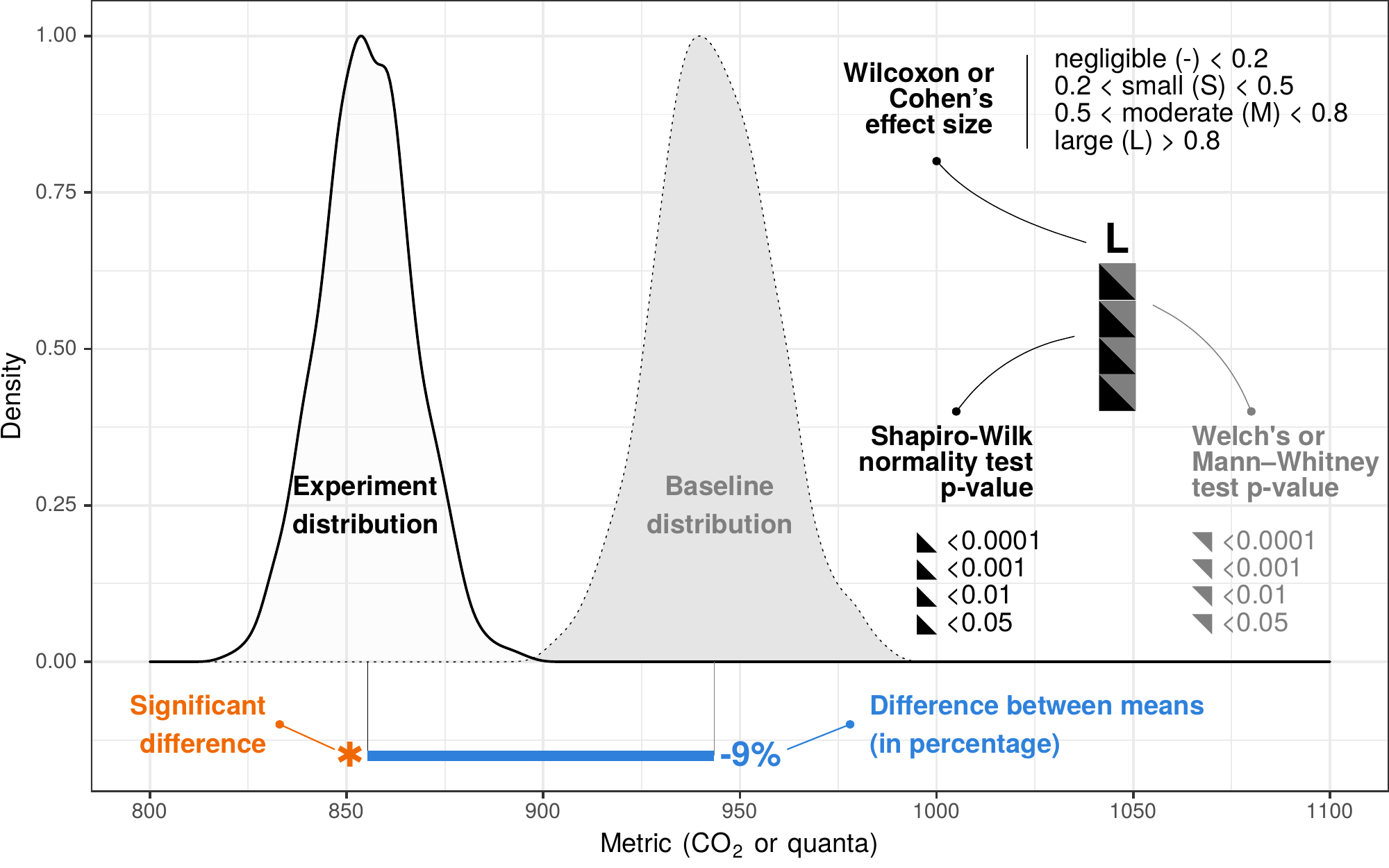}
    \caption{Statistical comparison of two distributions (baseline vs experiment). The legend is annotated to aid in the interpretation of subsequent figures.}
    \label{fig:legend_density}
\end{figure}

Each experimental configuration is compared to the baseline experiment to determine whether significant differences were achieved for all four outcome parameters. We hypothesized for each of the experimental configurations that the respective measures would positively impact IAQ and physiologic response, i.e., would yield lower $CO_2$ and $quanta$ levels in places and lower inhaled $CO_2$ and $quanta$ amounts in people compared to baseline. To test this hypothesis, we first apply the Shapiro-Wilk test of normality to the respective distributions. Then, for normally distributed data, Welch's t-test is used to compute a p-value, and the effect size \cite{Secchi2016} is calculated using Cohen's $d$. For not normally distributed data, Mann-Whitney U-test is used to compute a p-value, and the Wilcoxon test is used to calculate the effect size. Finally, the percentage difference between means is reported. A statistically significant difference between distributions, however, is only assumed whenever the p-value is below 0.001 \textit{and} the effect size is moderate (between 0.5 and 0.8) or large (greater than 0.8). \cref{fig:legend_density} provides a visual summary of the statistical analyses performed. 
All statistical tests are performed using R v3.6.3 \cite{R}. 

Note that for all quanta-related outcome parameters calculated in places, all simulation runs that yield a final $quanta$ level of zero at the end of the day are excluded from all experiments. The reason is that otherwise, the $quanta$ level distributions end up being bi-modal, with a skew towards zero. This happens predominantly for the chief offices if no infected person ever enters the office throughout the day. Having run each experimental configuration for many runs, we assume this to occur in the same frequency throughout all experiments - including the baseline experiment - which makes each experiment comparable to baseline when removing these simulations. In this way, only relevant simulations in which $quanta$ levels rose due to an infected being present are considered.



\section{Results and Discussion}

\subsection{Validation experiment}\label{validation}

\begin{figure}[!htb]
    \centering
    \includegraphics[width=0.55\linewidth]{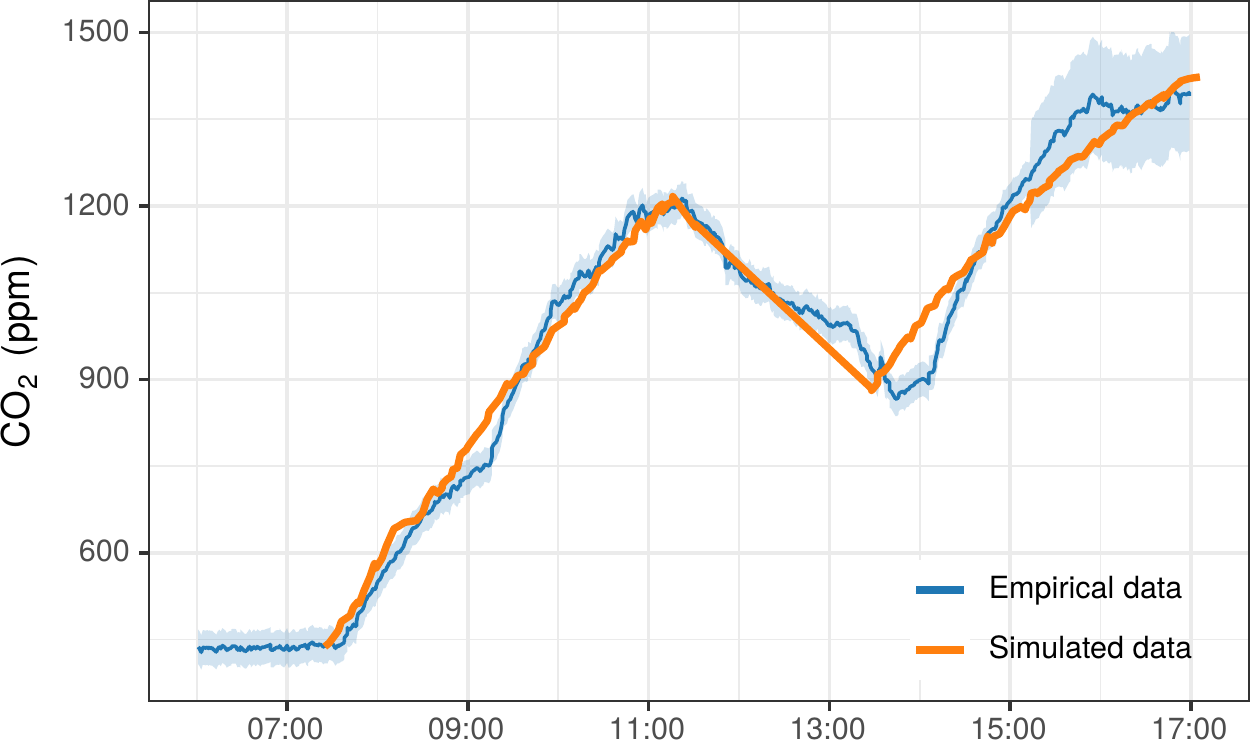}
    \caption{ArchABM validation experiment with empirical data registered by Candanedo et al. \cite{Candanedo2016} using a  
    Telaire 6613 sensor with range 0-2000 ppm and 1 ppm resolution. The blue shadow represents the accuracy of the sensor: $\pm 30$ ppm at 400-1250 ppm and $5 \%$ of reading $\pm 30$ ppm at 1250-2000.
    }
    \label{fig:validation}
\end{figure}

Recently, efforts have been made to analyze $CO_2$ levels inside buildings, and many studies can be found in the literature \cite{DiGilio2021,Jia2019,Candanedo2016}. 
The study of Candanedo et al. \cite{Candanedo2016}, presents measurements\footnote{\url{https://github.com/LuisM78/Occupancy-detection-data}} of $CO_2$ during a day in an office with two people present. Replicating their presented parameters for office room area and volume, timetable, and people present, we compare their data, which is available, to the results produced by ArchABM.
This comparison is presented in \cref{fig:validation}. In the real measurements, the $CO_2$ level begins to rise shortly before 8:00 a.m., when the room is first occupied. When the second occupant arrives just past 9:00 a.m., the slope of the $CO_2$ readings increases. Between 11:00 a.m. to 1:00 pm natural ventilation increases, obtaining outdoor air supply through door opening and/or occupants leaving, and thus the $CO_2$ decreases. Also, when the room is not occupied around 1:00 - 1:30 p.m., the $CO_2$ sensors register a slight drop in their readings. For this simulation, a constant outdoor air exchange rate of 0.25 $h^{-1}$ was set. It can be observed that the simulated data follow the measurements presented by Candanedo et al. \cite{Candanedo2016} in a satisfactory manner, and as a result, we proceeded with the experiments proposed in \cref{experiments}.

\subsection{Results for baseline experiment}

\begin{figure*}[!htb]
    \centering
    \subfloat[Density distribution for each activity throughout the day]{
    \includegraphics[width=0.5\linewidth]{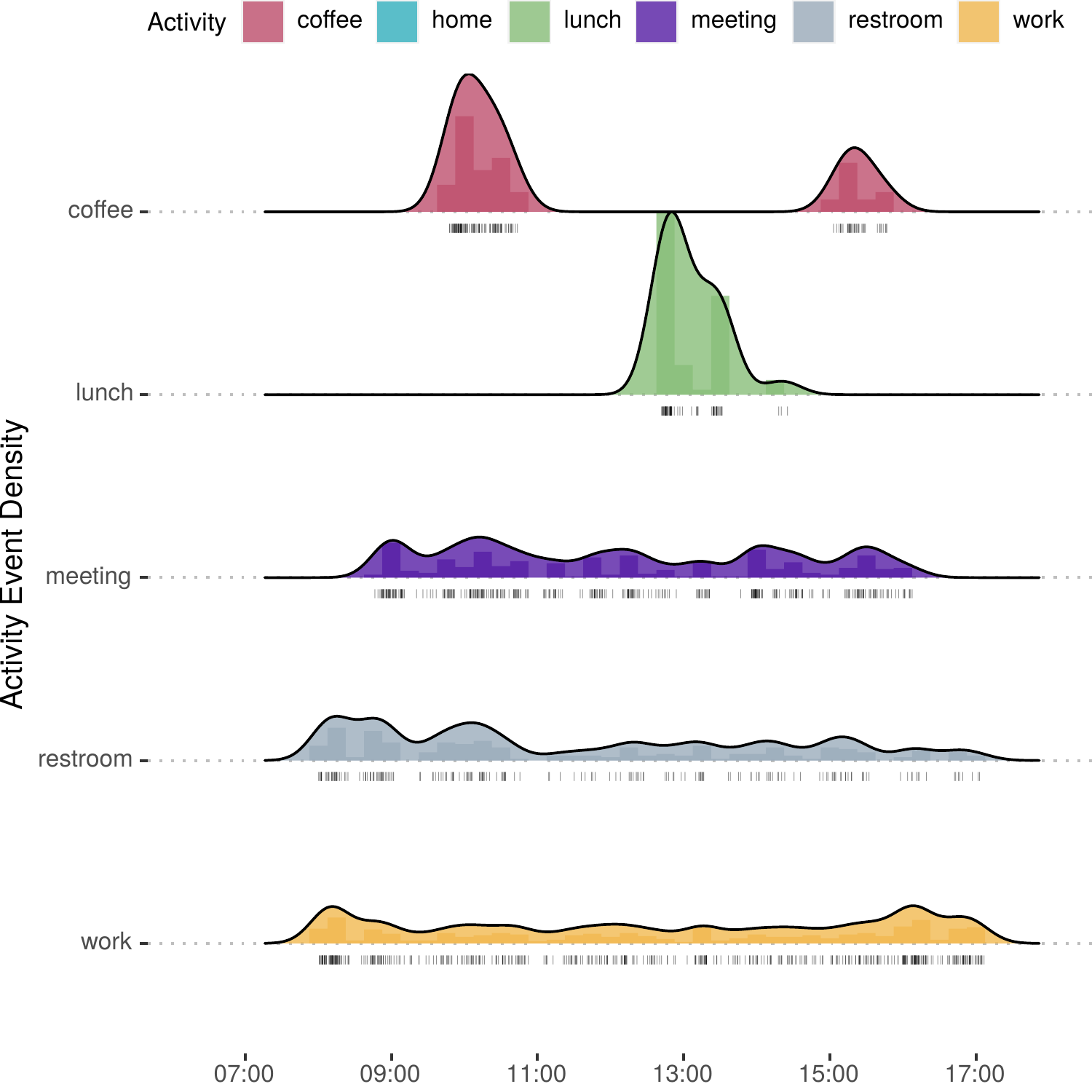}
    \label{fig:timeline_activity_density}} 
    \subfloat[Activities performed by each person throughout the day]{
    \includegraphics[width=0.5\linewidth]{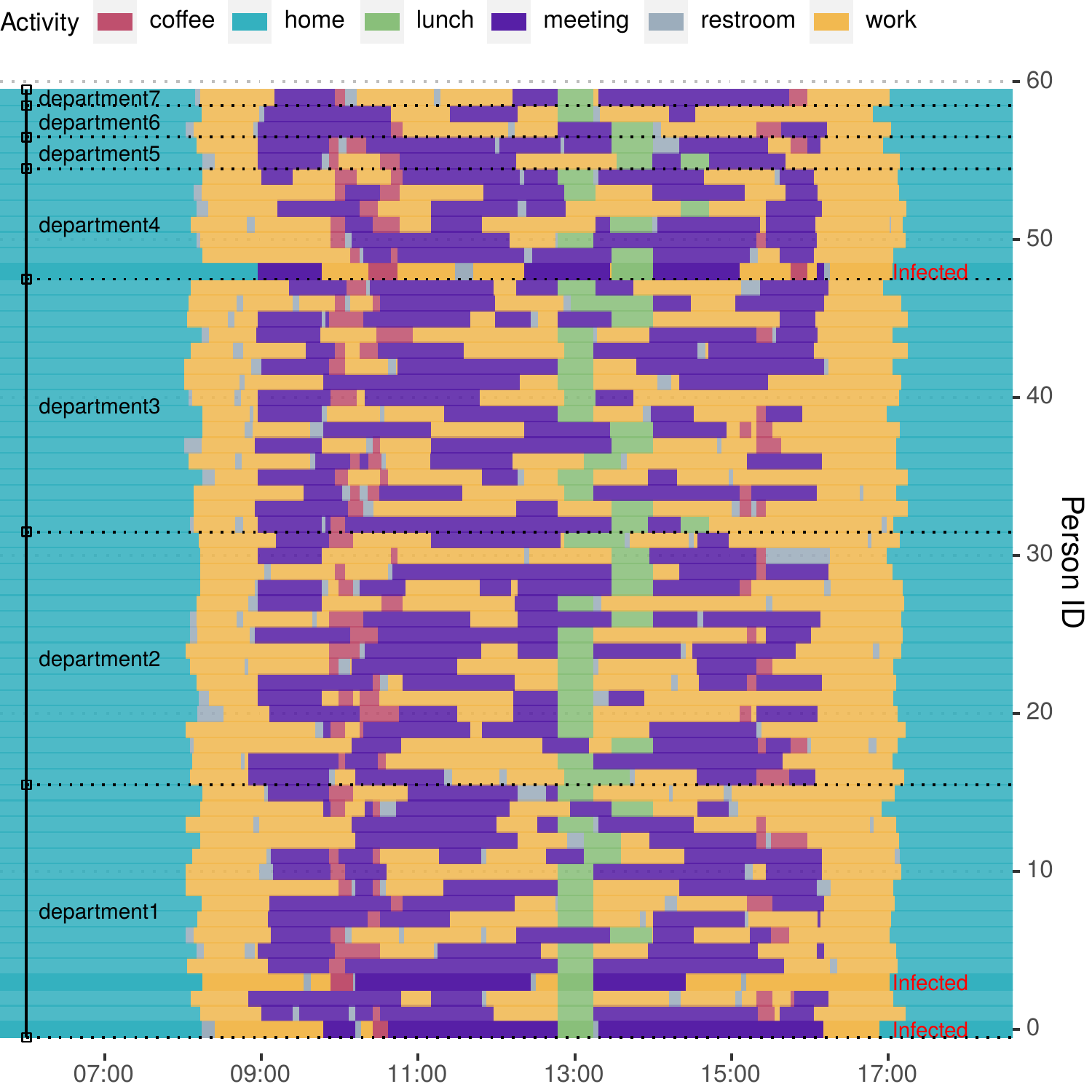}
    \label{fig:timeline_activity_person}} \\
    \subfloat[Quanta inhaled per person throughout the day. Infected people are marked with a dotted red line.]{
    \includegraphics[width=0.5\linewidth]{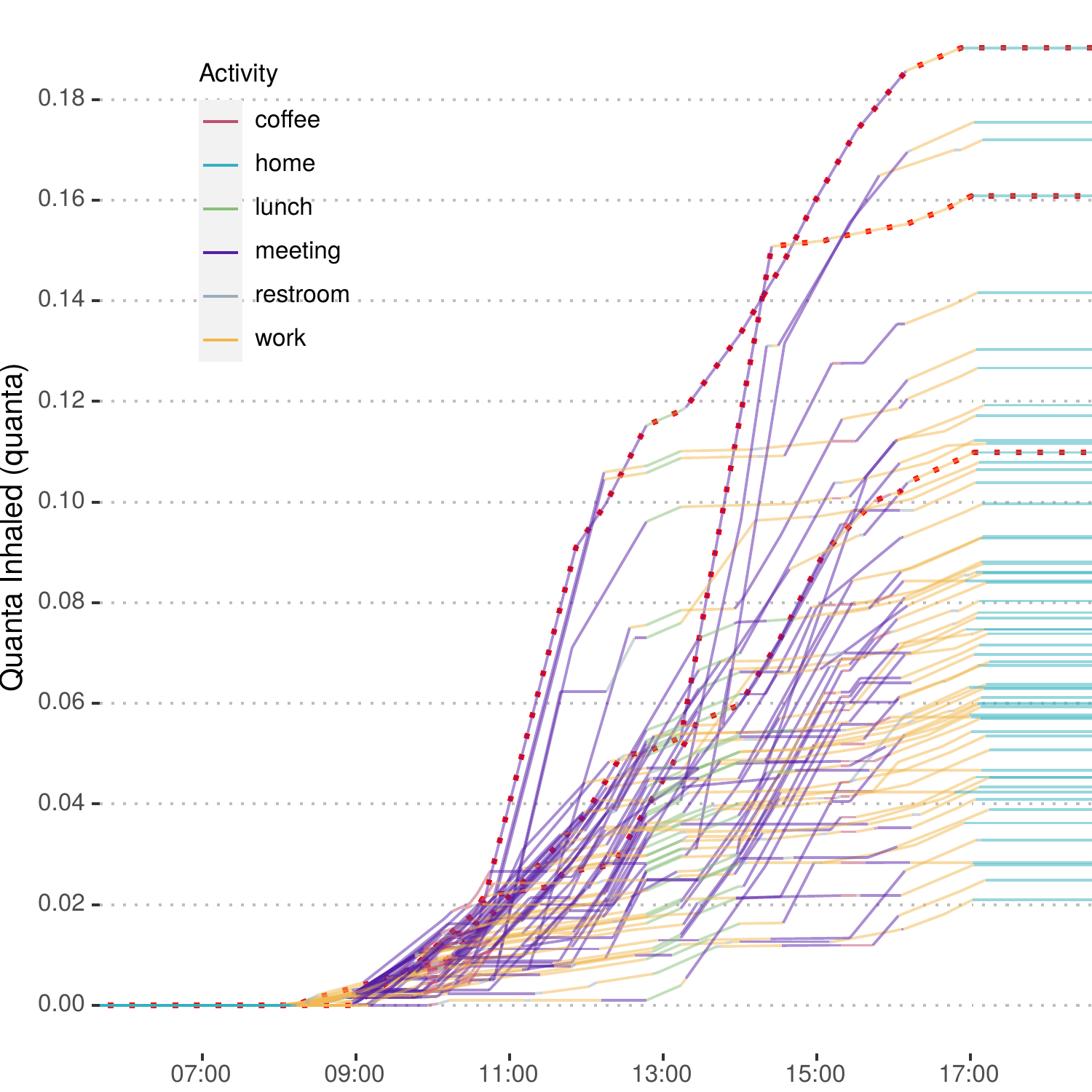}
    \label{fig:timeline_person_quanta}} 
    \subfloat[Quanta inhaled by each person at the end of the day.  Infected people are indicated with red dots.]{
    \includegraphics[width=0.49\linewidth]{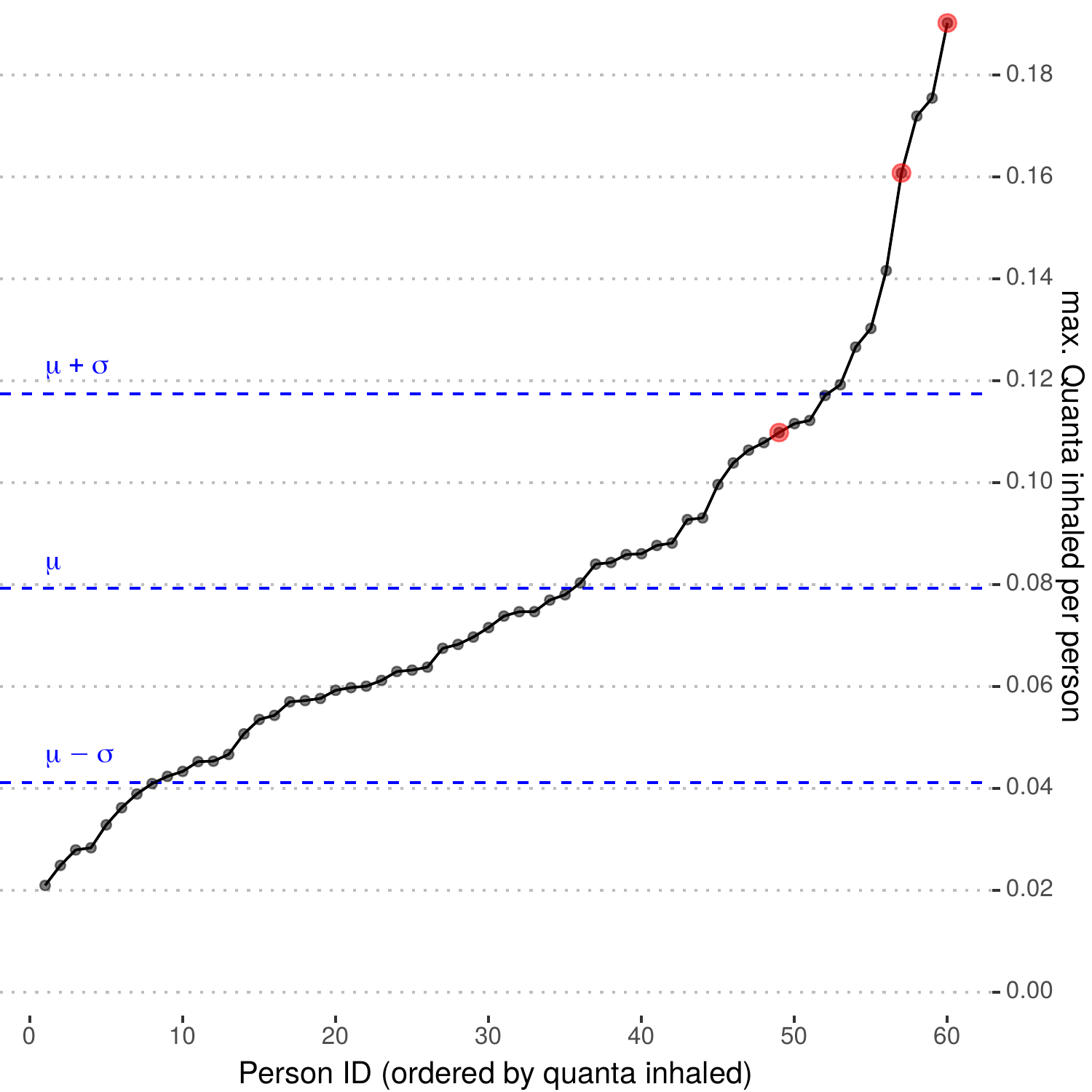}
    \label{fig:distribution_person_quanta}} 
    \caption{Baseline experiment results: activities throughout the day and $quanta$ inhaled per person}
    \label{fig:timeline_activity}
\end{figure*}

The results of a single simulation with the baseline configuration are presented in this section. \cref{fig:timeline_activity_density} summarizes the types of events (coffee, lunch, meetings, go to the restroom, do office work) performed by all occupants throughout the day, while \cref{fig:timeline_activity_person} shows a detailed breakdown of the activities performed by each person throughout the day. 

As it can be observed on \cref{fig:timeline_activity_person}, agents are strictly adhering to the specified schedule, with two coffee breaks, one main lunch event, and meetings, restrooms, and work events spread throughout the day. The three randomly infected people are also highlighted in \cref{fig:timeline_activity_person}.
The amount of quanta inhaled per person is depicted in \cref{fig:timeline_person_quanta}. Each line represents a person, and the red dotted lines indicate the three infected people. The color of the line represents the activity that each agent is performing. For instance, meetings and lunch activities primarily contribute to quanta inhalation between the agents. The total quanta inhaled by each person at the end of the day is shown in \cref{fig:distribution_person_quanta}, and the three infected people are highlighted with red dots.
\vspace{-6pt}

\begin{figure*}[!htb]
    \centering
    \subfloat[$CO_2$ level at each place throughout the day]{
    \includegraphics[width=0.4\linewidth]{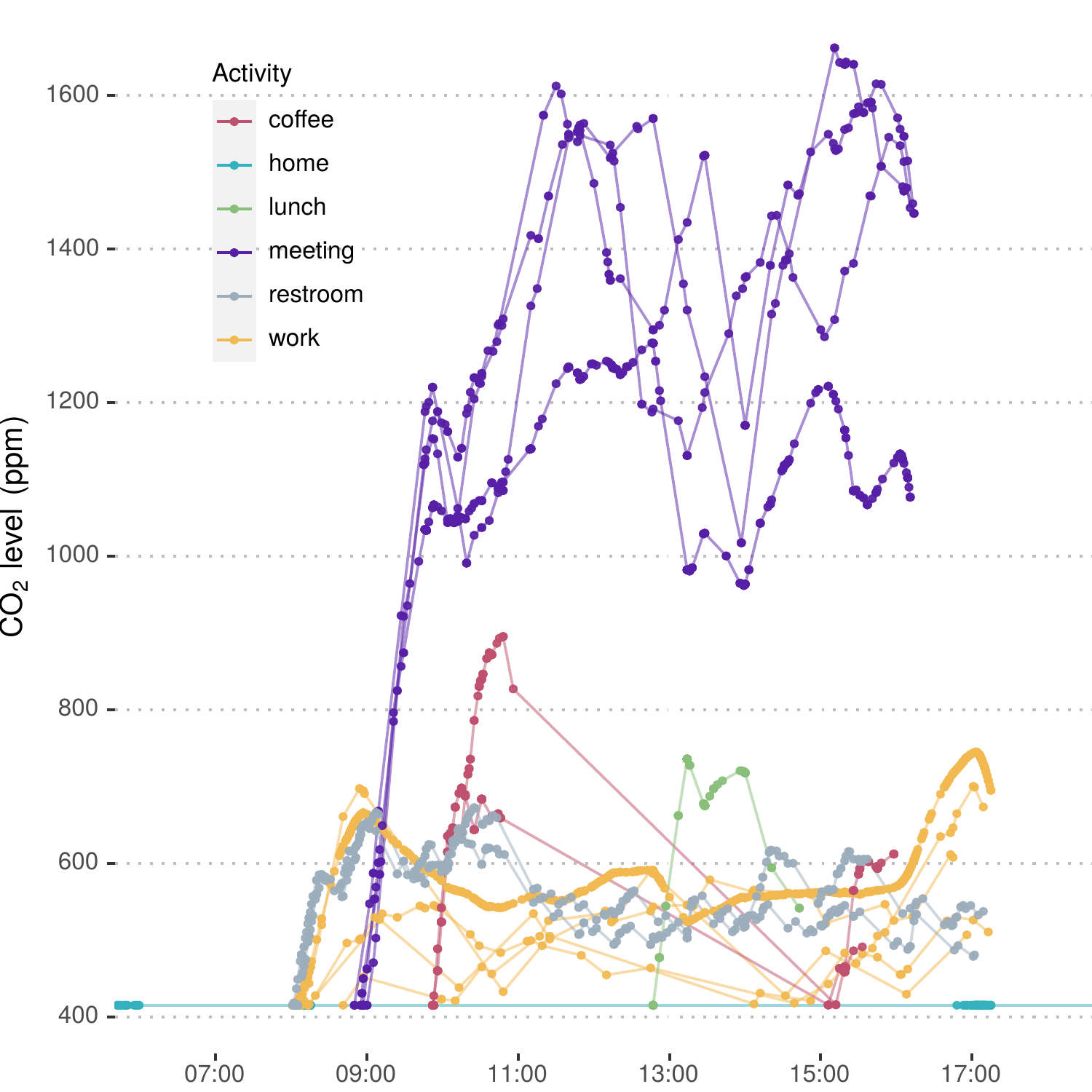}
    \label{fig:timeline_place_CO2}} 
    \subfloat[$CO_2$ level distribution at each place throughout the day]{
    \includegraphics[width=0.6\linewidth]{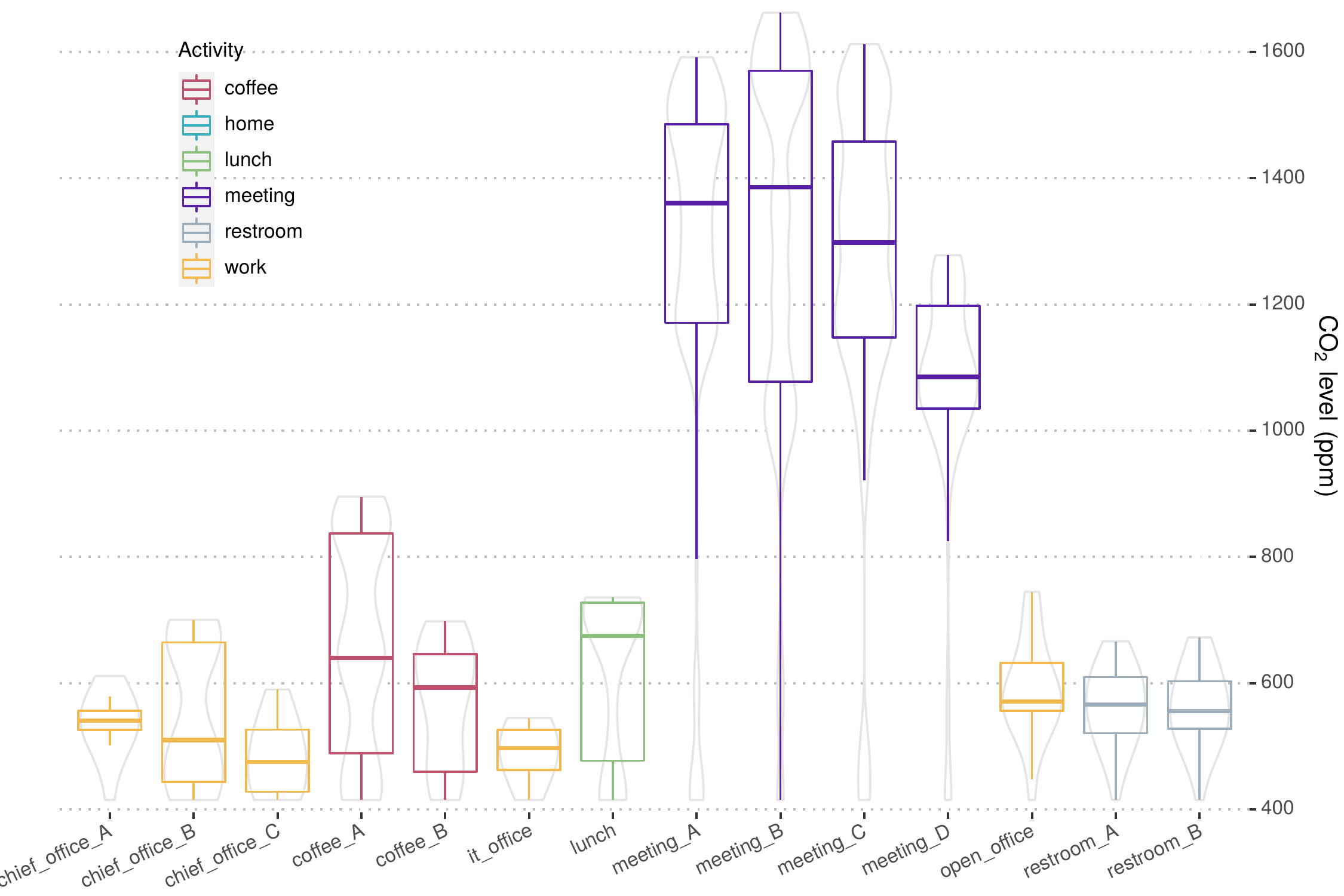}
    \label{fig:boxplot_place_CO2}} 
    \hfill
    \subfloat[$Quanta$ level at each place throughout the day]{
    \includegraphics[width=0.4\linewidth]{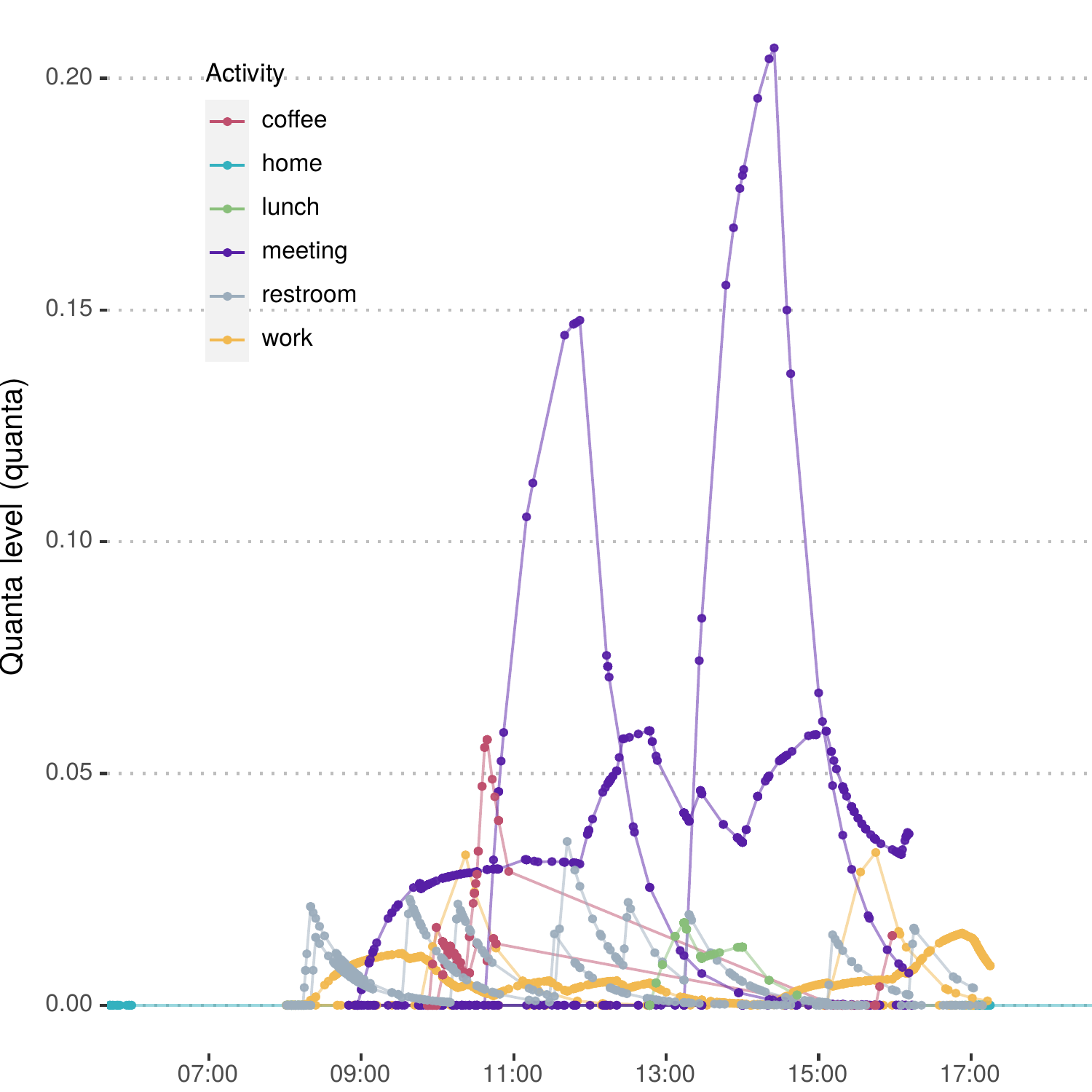}
    \label{fig:timeline_place_quanta}} 
    \subfloat[$Quanta$ level distribution at each place throughout the day]{
    \includegraphics[width=0.6\linewidth]{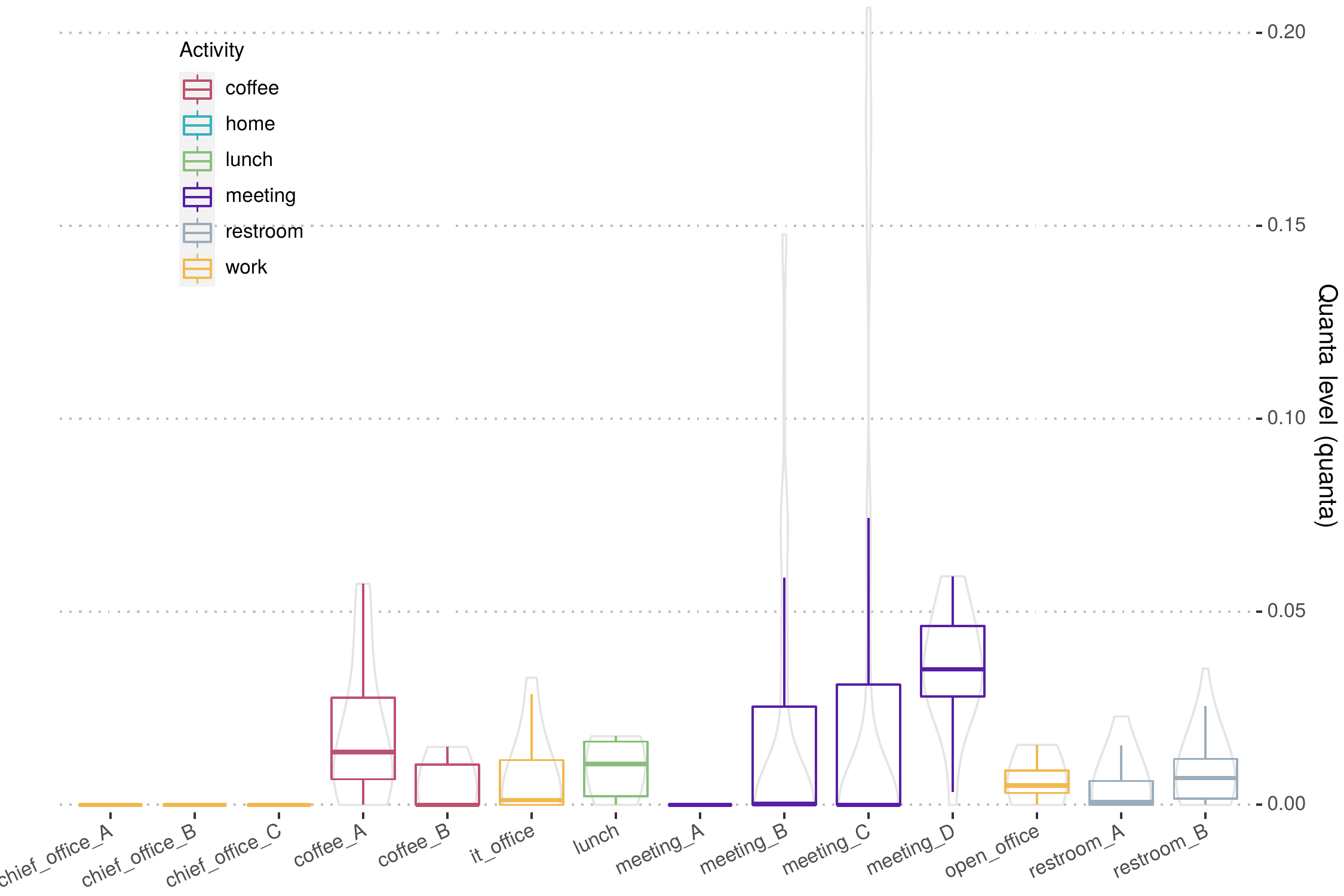}
    \label{fig:boxplot_place_quanta}} 
    \caption{Baseline experiment results: indoor air quality at place-level with $CO_2$ and $quanta$ levels}
    \label{fig:place_metrics}
\end{figure*}

From the places perspective, ArchABM also offers the possibility of tracking the $CO_2$ and $quanta$ concentration levels (\cref{fig:place_metrics}). Examining the $CO_2$ level at each place throughout the day (Fig. 10.a), it can be observed that the meeting rooms accumulate the highest $CO_2$ concentration throughout the day. The coffee places rapidly accumulate $CO_2$ during the coffee events, but the air quality is restored between the coffee breaks. Other rooms, for example, restrooms and office places show a more constant $CO_2$ level. The distribution of $CO_2$ concentration can directly be observed in \cref{fig:boxplot_place_CO2}, where a box-plot is overlaid on top of a violin plot. A similar interpretation can be concluded with the $quanta$ concentration for this simulation run. 

\begin{figure*}[!htb]
    \centering
    \subfloat[Maximum $CO_2$ level at each place]{
    \includegraphics[width=0.5\linewidth]{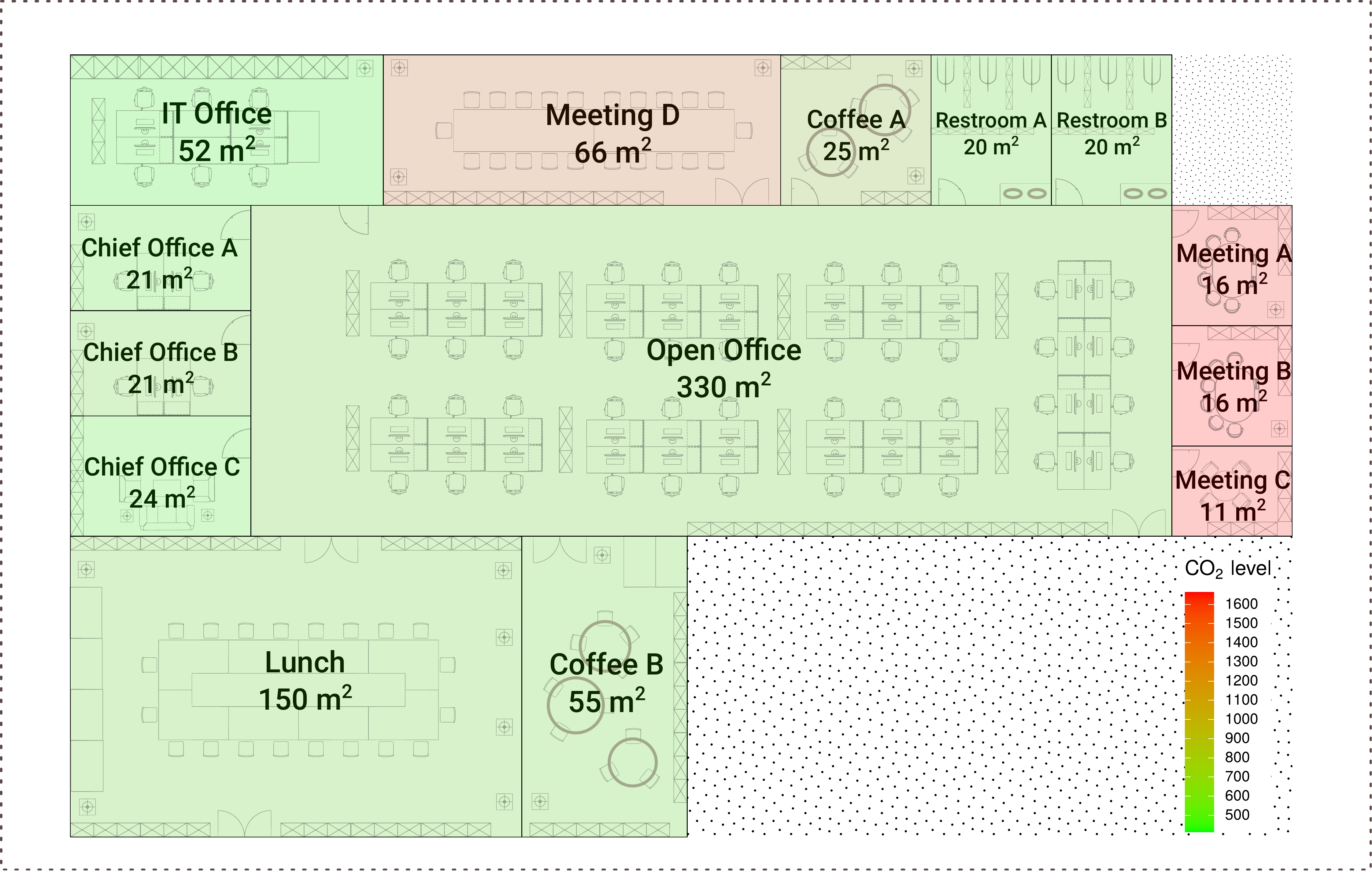}
    \label{fig:floorplan_CO2}} 
    \subfloat[Maximum $quanta$ level at each place]{
    \includegraphics[width=0.5\linewidth]{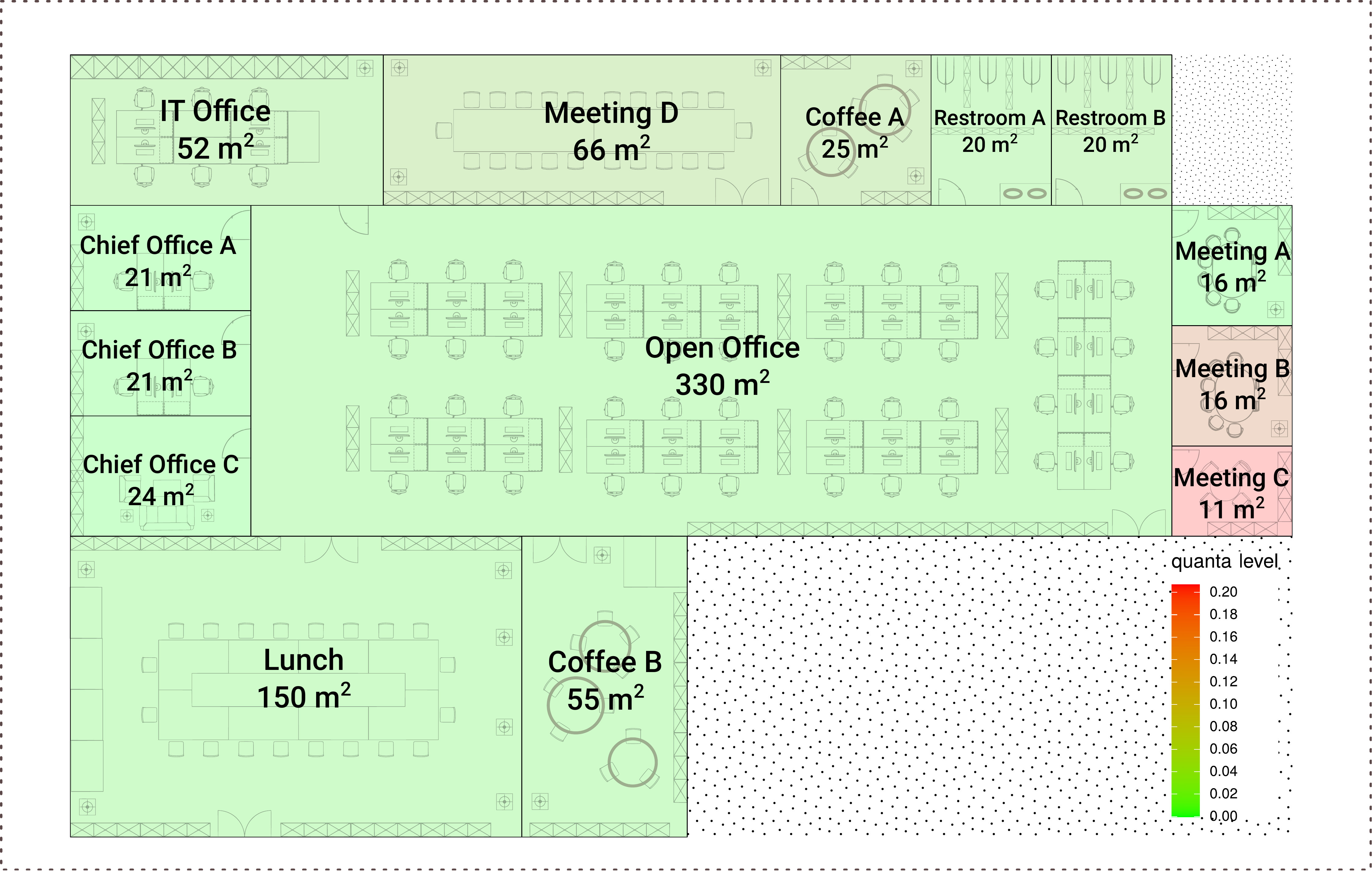}
    \label{fig:floorplan_quanta}} 
    \caption{Baseline experiment metrics related to IAQ at place-level overlaid on the floorplan}
    \label{fig:floorplan_results}
\end{figure*}

In addition, metrics related to IAQ at the place level have been overlaid on the floor plan, as shown in  \cref{fig:floorplan_results}. Concerning the $CO_2$ level, meeting rooms are highlighted as the worst locations. With regard to the $quanta$ level, meeting rooms B and C come out worst in this case. 
These results demonstrate ArchABM's capabilities of detecting "hotspots" in terms of high $CO_2$ and virus quanta concentrations (in our case meeting rooms and the coffee place) across the entire building.

It should be noted that the results in this section refer to a single simulation run and that the $quanta$-related metrics are very dependent on the randomly selected infected people. However, the high computational performance of ArchABM allows running multiple simulations, as is explained in the following section.

\subsection{Results for further experiments}

The impact of different building-related and company policy-related measures are presented in this section. As it was explained in \cref{experiments}, nine experiments are proposed, including the baseline case and the combined building-policy case. The results of the described experiments are evaluated with regard to three levels: places, people (i.e., departments), and the entire building.

\subsubsection{Results for places}

\begin{figure*}[!htb]
    \centering
    \includegraphics[width=0.95\linewidth]{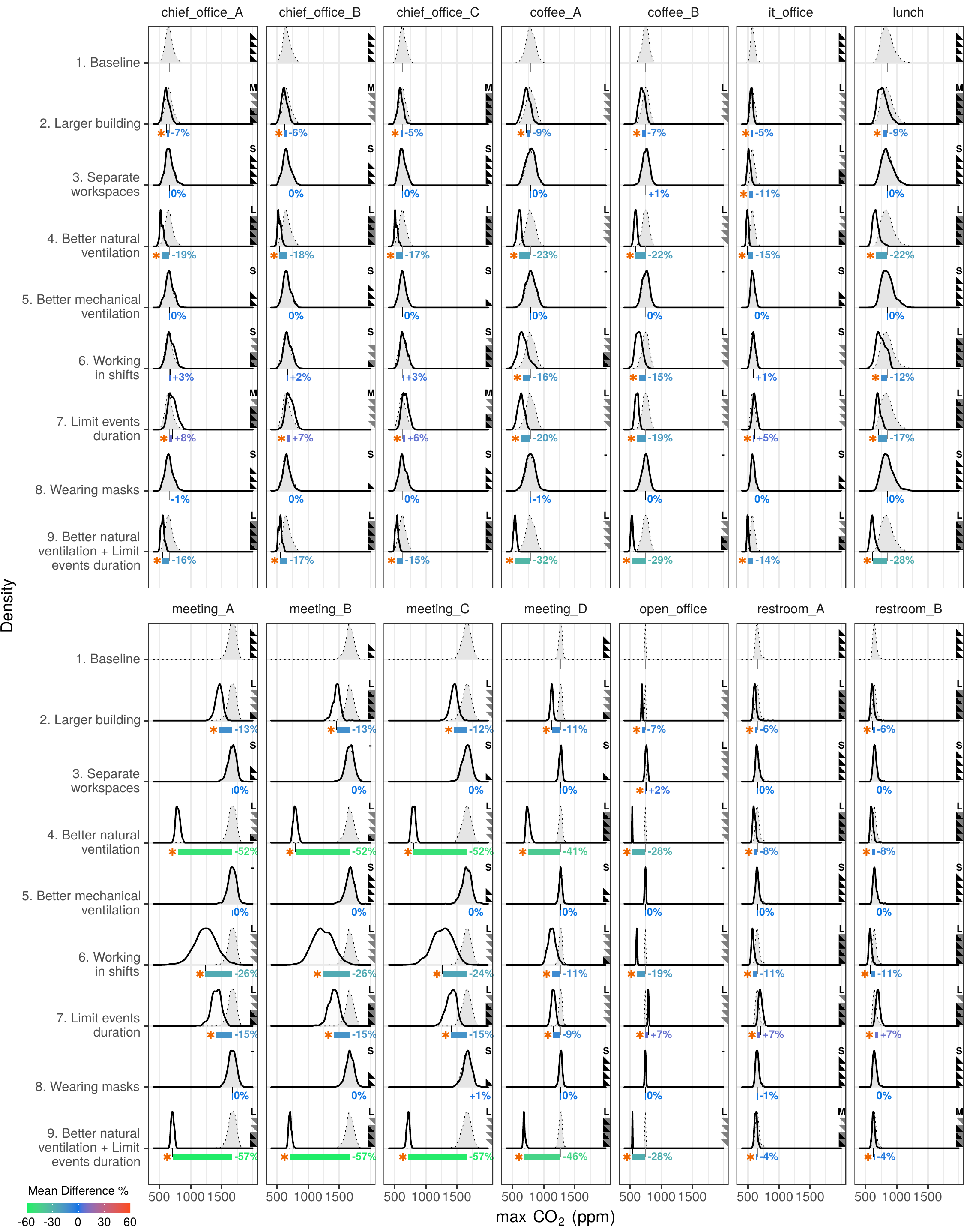}
    \caption{Indoor air quality at place-level: probability density function of the 
    maximum $CO_2$ level (concentration in ppm) at each place after $S_{run}=500$ simulation runs.}
    \label{fig:places_density_ridges_CO2}
\end{figure*}

\begin{figure*}[!htb]
    \centering
    \includegraphics[width=0.95\linewidth]{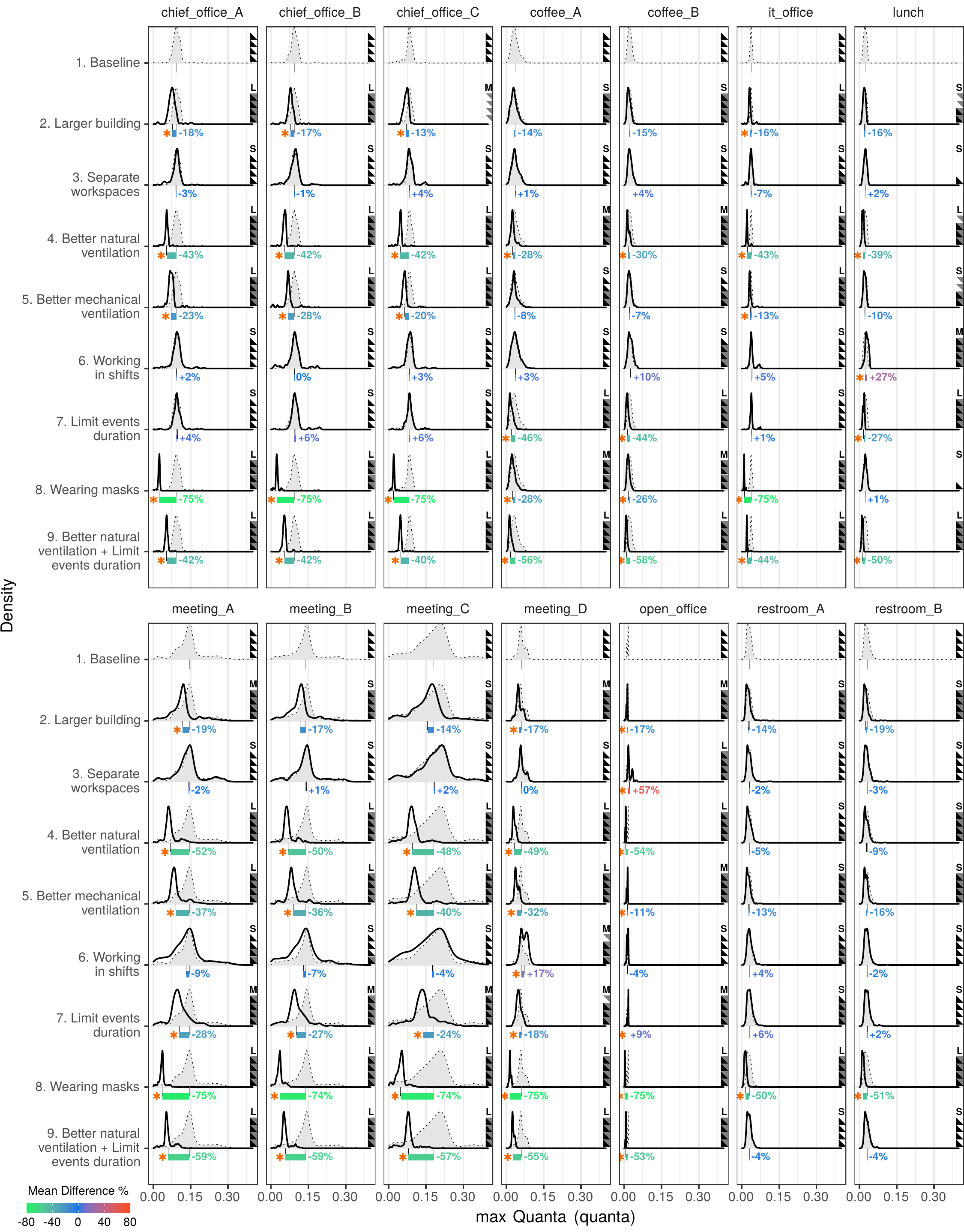}
    \caption{Indoor air quality at place-level: probability density function of the maximum virus $quanta$ level (concentration in quanta) at each place after $S_{run}=500$ simulation runs.}
    \label{fig:places_density_ridges_quanta}
\end{figure*}

\begin{figure*}[!htb]
    \centering
    \vspace*{-2cm}
    \includegraphics[width=0.86\linewidth]{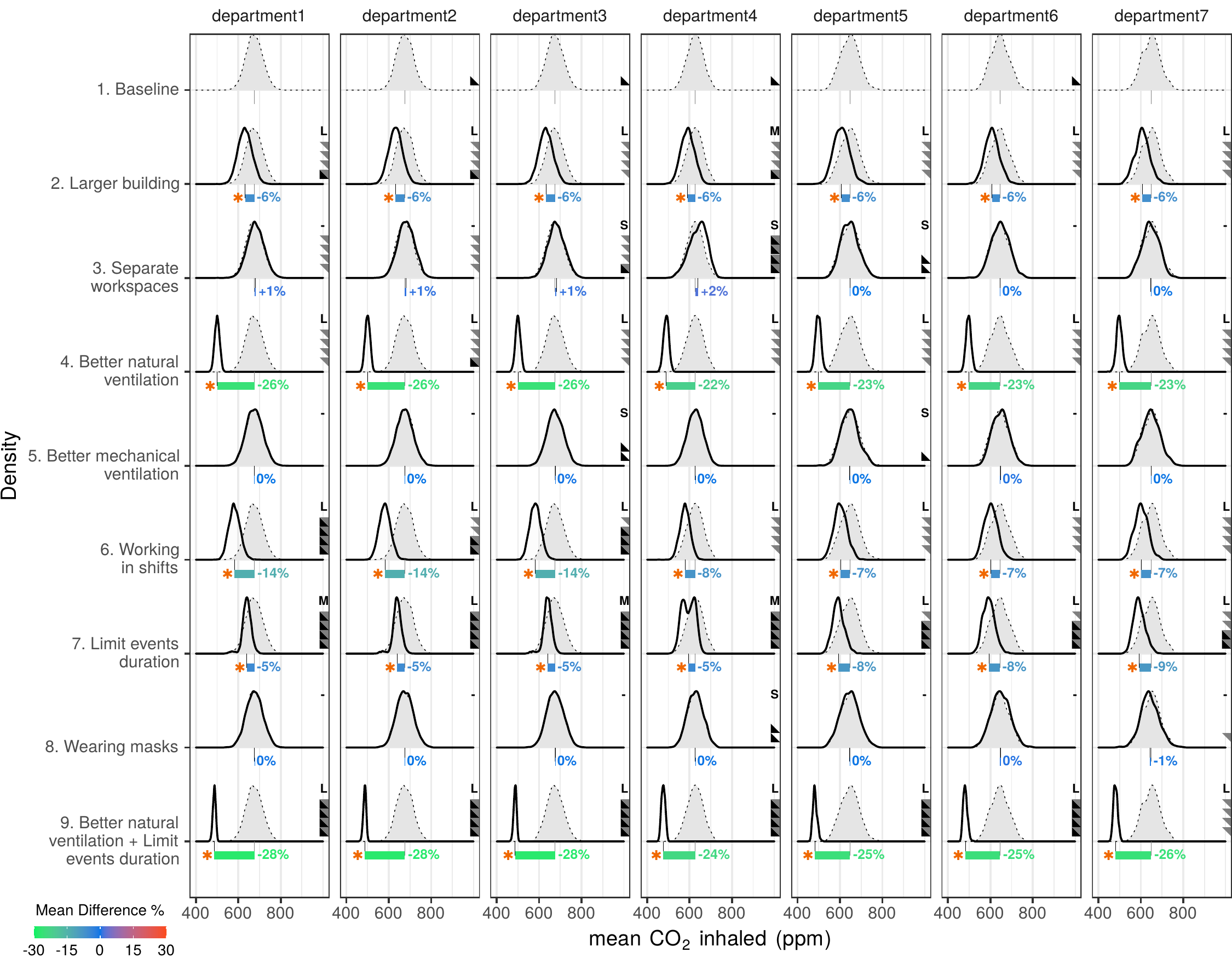} \\
    \includegraphics[width=0.86\linewidth]{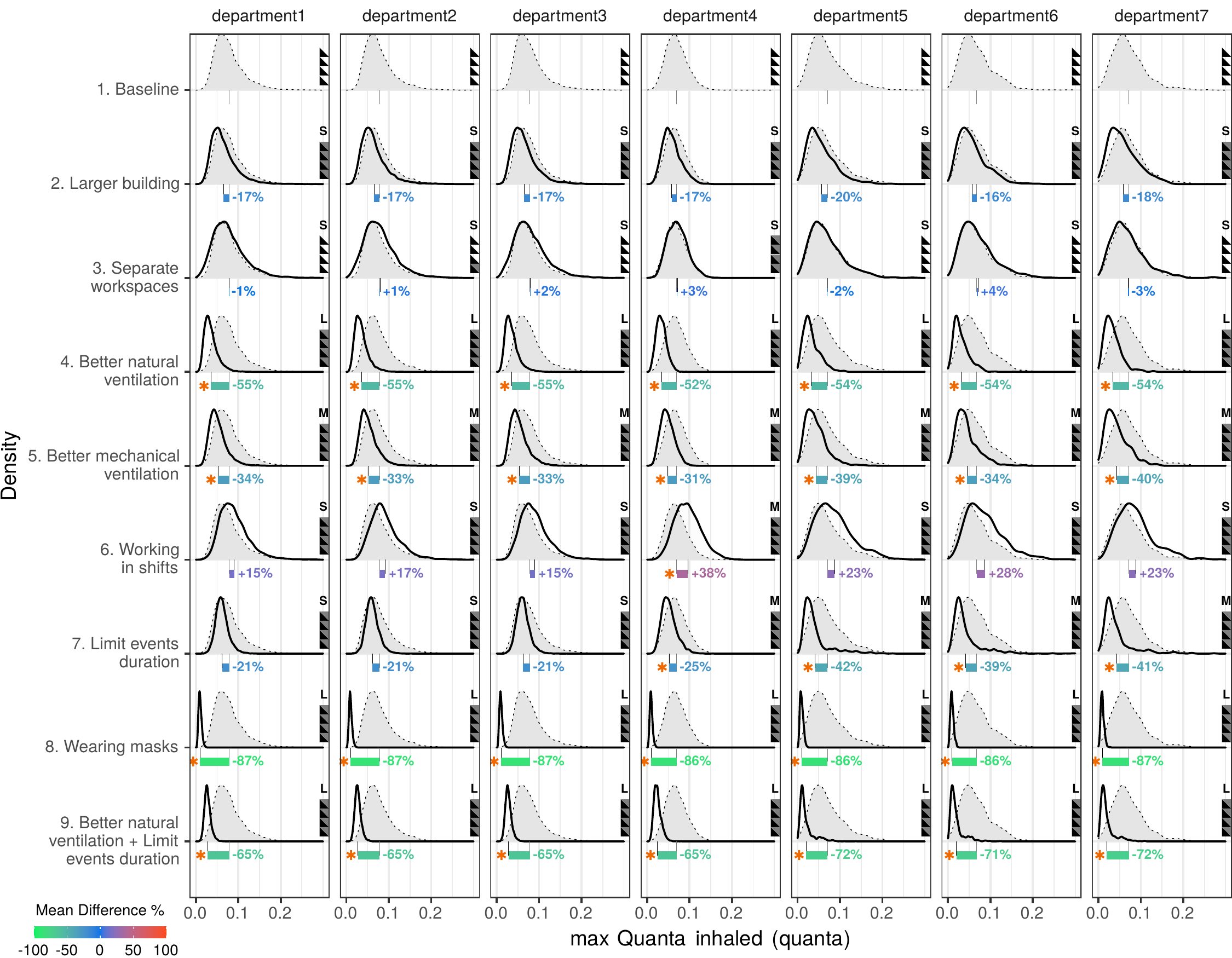}
    \caption{Physiological response outcome at person-level: probability density function of the time-weighted average inhaled $CO_2$ over the day and the maximum $quanta$ inhaled at the end of the day per person. }
    \label{fig:people_density_ridges}
\end{figure*}

In terms of outcome parameters related to IAQ at the place-level, the maximum $CO_2$ level (concentration in ppm) and the maximum virus $quanta$ level (concentration in ppm) reached during the day per place are calculated. This section examines the results from a location standpoint (\cref{fig:places_density_ridges_CO2} and \cref{fig:places_density_ridges_quanta}, and we will concentrate solely on differential aspects between places at each experiment. The analysis of these results takes into account the statistical significance of the hypothesis tests explained in \cref{statistical_analysis}.

\paragraph{Maximum $CO_2$ level} 

Designing a 20\% larger building (i.e. increasing each room's area by 20\%) reduces the maximum level of $CO_2$ in every room, and especially affects the meeting rooms, where the reduction nearly doubles when compared to the rest of the building. 
Separate workspaces have a significant impact only in the IT office and the open office. This strategy, in particular, lowers the maximum level of $CO_2$ in the IT office by up to 11\% while increasing it by 2\% in the open office.
Designing better natural ventilation systems greatly improves the indoor air quality in terms of $CO_2$ level, particularly in meeting rooms. Improved mechanical ventilation systems, on the other hand, have no effect on $CO_2$ levels due to the way they were implemented in the model (no air exchange/replacement). 

The strategy of working in shifts affects the entire building, except for the chiefs' rooms, and reduces the maximum $CO_2$ level of the meeting rooms in particular. 
Shortening the duration of events produces interesting results in terms of $CO_2$.
People are not allowed to take long coffee breaks, lunch activities or meetings, so they spend more time in other places (for instance, chiefs, IT office, open office and restrooms), where the $CO_2$ level rises.
Establishing a mandatory mask policy has no effect on the $CO_2$ level as the model does not take mask filtration into account for $CO_2$ calculations.
Finally, the combination of improved natural ventilation and events duration limitation corresponds to the confluence of the aforementioned observations about these experiments.

\paragraph{Maximum $quanta$ level}

The design of a larger building in terms of room area reduces the maximum $quanta$ level in every room by up to 18\%.
Separate workspaces have a significant impact exclusively in the open office, which is divided into three distinct spaces according to this strategy.
This building configuration specifically raises the maximum $quanta$ level in the open office by up to 57\%.
This increase in the mean $quanta$ level is due to the fact that in this experiment, one of the three spaces is more likely to be highly contaminated, which raises the mean value. 
Better natural ventilation system design improves indoor air quality in terms of $quanta$, especially in meeting rooms. Installing better mechanical ventilation systems reduces $quanta$ concentration levels in all rooms, with a greater impact in chief offices and meeting rooms.

The strategy of working in shifts increases the prevalence of the virus, affecting the $quanta$ levels of the entire building as we kept the number of infected people constant in our experiments. Larger spaces where a large number of people can congregate (such as the lunch room and the largest meeting room D) are particularly affected.  
Limiting the duration of events produces the same results as analyzing the $CO_2$ level under this strategy. People spend more time in other places (for example, chiefs, it office, open office, and restrooms, where the $quanta$ level increases) because they are not allowed to take long coffee breaks, lunch activities, or meetings.
Establishing a mandatory mask policy affects every area in the building, except the lunch room, and has a greater effect on the $quanta$ level of the chief offices and meeting rooms. Combining the experiment of improved natural ventilation with the limitation of events' duration produces very similar results to the former experiment. However, in this case there is a slight improvement in the quanta level in the coffee places and the lunch room.

\subsubsection{Results for people / departments}

In terms of physiological response outcome at person-level, the time-weighted average inhaled $CO_2$ over the day and the maximum $quanta$ inhaled at the end of the day per person are used. This section examines the results from a departmental viewpoint (\cref{fig:people_density_ridges}), and we will concentrate solely on differential aspects between departments at each experiment. Once again, the analysis of these results takes into account the statistical significance of the hypothesis tests explained in \cref{statistical_analysis}.

Working in shifts reduces the average of $CO_2$ level inhaled by all departments, but particularly in departments D1, D2, and D3. This difference is clearly explained by the fact that in this experiment the number of people in these departments gets reduced, and thus, the emitted $CO_2$ decreases as well. Nonetheless, in terms of $quanta$ level in this shift-work strategy, maintaining a constant number of infected people while reducing the total number of people increases the virus prevalence among the population, which has the greatest impact on the IT office department. As a consequence, the $quanta$ level at the department D4 (IT office) is increased by 38\%, while the rest of the departments suffer an increase of 15 to 25\%, approximately. 

Limiting the duration of events has a similar effect with respect to the average $CO_2$ level inhaled across departments, with an average reduction of 5 to 9 percent. However, in terms of maximum $quanta$ inhaled at the end of the day per person, this strategy has the greatest impact on departments D5, D6 and D7, which represent chief officers. In comparison to departments D1-D4, the reduction in average $quanta$ inhaled by departments D5-D7 is nearly doubled. Interestingly, other ABM-based studies found similar results when analysing the duration of classes in schools \cite{Zafarnejad2021}, where shorter classes were preferable. The remaining experiments (\textit{larger building}, \textit{separate workspaces}, \textit{better natural ventilation}, \textit{better mechanical ventilation}, and \textit{wearing masks}), do not show any differential aspects among departments, however, wearing masks has the largest impact on reducing quanta levels across all departments. The benefit of wearing masks indoors has been found in several other studies applying ABMs \cite{DOrazio2020, Farthing2021}.

\subsubsection{Results for whole building}

\begin{figure*}[!htb]
    \centering
    \includegraphics[width=0.65\linewidth]{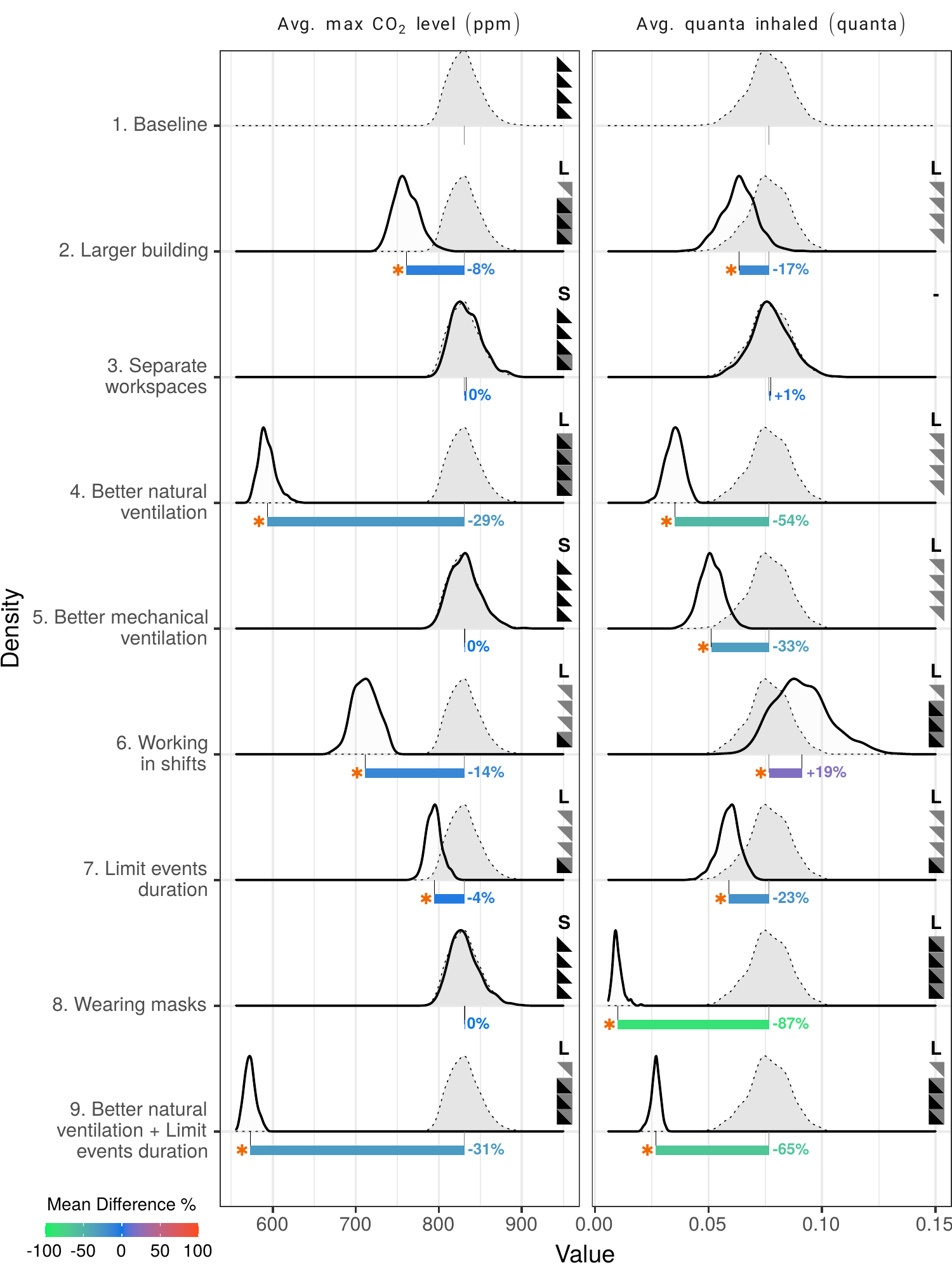}
    \caption{Results at building level: Indoor air quality is measured by the volume-weighted average maximum $CO_2$. Physiological response outcome is measured by the maximum $quanta$ level inhaled at the end of the day averaged over all people.}
    \label{fig:building_density_ridges}
\end{figure*}

At the building level, volume-weighted average maximum $CO_2$ is reported in terms of IAQ parameters per experiment, where volume refers to the volume of each place. To summarize physiological response parameters on the building level, the maximum $quanta$ level at the end of the day is averaged over all people. Once more, the analysis of these results takes into account the statistical significance of the hypothesis tests explained in \cref{statistical_analysis}, which has been included on \cref{fig:building_density_ridges}

Concerning the building-related measures, increasing each room's area by 20\% reduces, on average, the maximum $CO_2$ level by 8\% and the maximum $quanta$ level by 17\%. However, the cost of these solutions must be carefully considered, and in some cases, they are not a financially viable option. Creating separate workspaces does not affect either the $CO_2$ nor $quanta$ levels at the building level. However, the results from the perspective of the place claim that it affects the modified spaces. Increasing the natural ventilation, the outdoor air exchange rate, reduces, on average, the maximum $CO_2$ level by 29\% and the maximum $quanta$ level by 54\%. This measure improves the IAQ of the building and is a crucial parameter to control the indoor air quality, as expected. Increasing the mechanical ventilation rate improves the $quanta$ level by 33\% but does not modify the $CO_2$ concentration level, as there is no outdoor air supply, the air is merely recirculated. Although virus quanta can be removed from recirculated air, the $CO_2$ level remains unchanged. 

Regarding the company policy-related measures, working in shift reduces, on average, the maximum $CO_2$ level by 14\%, but increases the maximum $quanta$ level by 19\%. These results can be explained by the fact that maintaining a constant number of infected people while reducing the total number of people increases the virus prevalence among the population. Thus, it is expected to observe the $quanta$ level increase in this experiment. Constraining the events' duration reduces, on average, the maximum $CO_2$ level by 4\% and the maximum $quanta$ level by 23\%. This reflects the fact that the $quanta$ level is directly related to the meeting events and rooms and that limiting their duration can be an easy-to-apply and effective strategy to reduce the $quanta$ concentration indoors. Establishing a mandatory mask policy has no effect on the $CO_2$ level but almost completely eliminates the $quanta$ concentration from the building, as predicted by the aerosol model developed by Peng and Jimenez \cite{Peng}. Other ABM models from the literature, such as \cite{Antczak2021,Farthing2021,Altamimi2021,Zafarnejad2021}, have made similar observations regarding the use of masks.

Combining better natural ventilation and limiting the duration of meetings and lunch events has a significant effect on both $CO_2$ and $quanta$ levels. This case combines the most promising measures from the above experiments and reduces, on average, the maximum $CO_2$ level by 31\% and the maximum $quanta$ level by 65\%.













\subsection{Strengths, limitations and future work}

In this study, we present a novel agent-based simulator framework, ArchABM, designed to model complex human-building interaction patterns for evaluating the impact of building-related and policy-related measures on a building's IAQ in terms of $CO_2$ and virus quanta levels and the respective physiological response of occupants. While the application of ABMs in the building domain has grown in recent years, few studies specifically focus on IAQ, and only very recent studies consider the latest call to include potential airborne virus transmission in their models. Furthermore, to the best of our knowledge, no other ABM-based approach has been published that allows for an easy-to-setup simulation of complex, realistic human-building interaction patterns resulting from occupants following their daily schedule across an entire building and adhering to conventions or local policies while at the same time including a random component.

When designing a building or adapting an existing building to improve IAQ for its occupants on a broad scale, cost and energy efficiency are critical considerations. For example, a large, perfectly ventilated room may not be necessary depending on the given human interaction with this room. At the same time, it may be futile to optimize one room's ventilation system under great (engineering and monetary) efforts when occupants later get infected in another, overlooked room such as a coffee kitchen. ArchABM was designed to provide a general overview and optimize the entire building based on its intended use. Simulations performed with ArchABM can assist in unveiling such "hotspots" with poor IAQ or "oversized" rooms and allows playing with a combination of (likely costly) building-related measures and (potentially less expensive) policy-related measures that may lead to significantly improved IAQ. Thanks to ArchABM's good computational performance, results can be evaluated quickly and with a high number of simulation repetitions, which allows extracting statistically valid conclusions for decision-makers.

We present here a realistic use case applying ArchABM, based on a real floor plan with realistic room sizes, number of people, and workday routines, which results in a series of interaction patterns that in turn lead to different IAQ distributions in different rooms. While some of the investigated measures resulted in expected outcomes in terms of $CO_2$ and virus quanta levels (such as masks having a positive effect on quanta levels and inhaled quanta), ArchABM allows quantifying and comparing these effects in detail - revealing, for example, that relatively "simple" measures such as improved natural ventilation (increase outdoor air supply by opening windows) or reducing the duration of events have a relatively significant positive impact. Results from our simulations further show how poor IAQ in terms of $CO_2$ and virus quanta levels can be detected for each room in the building, highlighting in our case that in particular meeting rooms are problematic in terms of IAQ. ArchABM allows combining, testing and quantifying a set of various measures, whose outcome may not be easy to predict, taking into account the use of the entire building by its occupants. Combining better natural ventilation with limiting meeting duration, for instance, could decrease average $CO_2$ levels on the building level by 31\% and average virus quanta levels by 68\%. In those scenarios, ArchABM can provide valuable quantitative insights to architects, engineers, or building and human resource managers.

\subsubsection{IAQ including aerosols}
IAQ in this study has been principally defined by pollutant concentration. A recently published aerosol model calculating both the amount of exhaled and inhaled $CO_2$, as well as virus quanta (here for SARS-CoV-2), was adapted to account for the dynamics of people entering and leaving different rooms over a day to be applicable in our agent-based simulation. ArchABM's calculations regarding $CO_2$ levels in a room were validated against a well-documented case from the literature where $CO_2$ was measured, and recorded \cite{Candanedo2016}. Validating inhaled $CO_2$ and in particular accumulated and inhaled virus quanta levels is to date as complex as to measure virus concentrations in a room reliably. Regarding SARS-CoV-2, this is currently a critical research topic, and viable options may become available in the future. For now, we have to rely on the applied aerosol model, which in any case has been validated against real cases of SARS-CoV-2 transmissions \cite{Miller2021}. We would like to point out that here we deliberately refrained from calculating parameters such as "infection risk" as we understand that it is still not fully known which amount of virus quanta leads to which amount of risk. We therefore only report virus quanta levels for rooms and inhaled quanta for occupants.
ArchABM being flexible and modular, and available as an Open Source Python library, it is straightforward to adapt the current SARS-CoV-2 aerosol model or another model related to airborne virus transmission. The same applies to extending the definition of the calculated IAQ in order to include other relevant parameters such as temperature, or humidity \cite{Candanedo2016} - this can be addressed in future work.

\subsubsection{Localised air flow}
Compared to other ABM-related studies using aerosol models for indoor environments, to the best of our knowledge, this study is the first to analyse differences between mechanical and natural ventilation configurations which apply varying ventilation rates. One limitation of the aerosol model included in ArchABM though is related to the nature of ventilation, which removes both $CO_2$ and virus quanta from a room over time. Here, we assume homogeneous mixing of air constituents and do not take into account airflow direction or any localized airflows induced either by people or the ventilation itself, depending on the room geometry or interference with objects or occupants in the room. This is currently another important research topic, with authors conducting detailed (mostly Computational Fluid Dynamics (CFD)-based) studies of indoor airflows and different ventilation systems \cite{Vuorinen2020,Pang,Bhagat2020,Barbosa2021,Shao2021}. Evidence suggests that the position of ventilation inlets and outlets, as well as occupants' individual positions relative to an infected person play an important role in airborne virus transmission \cite{Barbosa2021,Zafarnejad2021}. Accounting for these complex air flow-related effects is generally difficult in agent-based models as computational efforts even for a single room are typically high. However, recent studies suggest model adjustments taking into account airflow directions or locally spread quanta \cite{Altamimi2021,Zafarnejad2021}. 

\subsubsection{Agent characteristics}
Regarding the agents' (i.e., building occupants') characteristics used in this study, future work could involve varying agents' profiles in order to account for different types of respiratory activities (speaking, shouting, breathing, etc.), respiratory parameters (e.g., inhalation rate) or activity level (resting, standing, walking, etc.) \cite{Buonanno2020}. Even varying individual agent behavior based on agent physical perception or physiology may be added \cite{Jia2019,Tijani2016}. Yet, adding such complexities to the model will result in performance loss and may reduce the comprehensibility and explainability of the model's output. For certain applications, such a level of detail regarding "agent microdynamics" may not be necessary.

\subsubsection{Building optimisation}
ArchABM's good computational performance and standard output format (JavaScript Object Notation, JSON) make it attractive for coupling with advanced optimization methods such as Reinforcement Learning \cite{Valladares2019} or more traditional Knapsack \cite{demirovic2019investigation}, or Design of Experiments approaches \cite{Thiele2014}, with which optimal room sizes and distributions or ventilation configurations could be estimated based on simulated human-building behavior. We believe there is an immense potential for future studies to explore such building design optimization in order to find cost and energy-effective solutions that at the same time provide good IAQ across the entire building - given its use.

\section{Conclusion and summary}
In this study, we present a novel, fast and open source agent-based modelling framework, ArchABM, which allows for simulating complex human-building-interaction patterns to estimate IAQ across the entire building, while taking into account potential airborne virus concentrations. A recently published aerosol model for SARS-CoV-2 was adapted to calculate time-dependent carbon dioxide ($CO2$) and virus quanta concentrations in each room as a measure of IAQ as well as inhaled $CO2$ and virus quanta for each agent (occupant) over the course of a day as a measure of physiological response. ArchABM was then applied to simulate a realistic office scenario including 14 rooms and 60 agents to investigate the impact of building-related measures and policy-related measures on overall IAQ and physiological response of occupants. Results allowed determining critical and not-so-critical rooms in terms of IAQ and allowed for a quantitative assessment of the impact of single and combinations of measures on IAQ and physiological response, suggesting that improved natural ventilation, limiting meeting duration and wearing masks were among the most effective measures.

The pandemic caused by SARS-CoV-2 has demonstrated that we need to focus more on improving IAQ in order to avoid risking occupants' health. We believe that advanced simulation tools such as ArchABM can provide novel insights and moreover may assist stakeholders in finding cost and energy effective solutions, ensuring good IAQ across a building, while taking into account how single rooms are actually used. Using ArchABM in optimisation scenarios may even lead to new ways of designing future buildings that provide healthier indoor environments.

\section*{Declaration of competing interest}
The authors declare that they have no known competing financial interests or personal relationships that could have appeared to influence the work reported in this paper.

\section*{Acknowledgements}
The authors would like to thank Asier Aguirre Martínez and Egoitz Bizkarguenaga Agirre for their help in obtaining realistic room sizes, floor plans and work schedules for the use case of our research center. The authors further thank the anonymous Reviewers for their insightful comments that helped to improve the manuscript.

\section*{Copyright Information}
\noindent
\begin{minipage}[t][][c]{0.8\textwidth}
This manuscript version is made available under the CC BY-NC-ND 4.0 license \\
\url{https://creativecommons.org/licenses/by-nc-nd/4.0/deed.en}
\end{minipage}
\hfill
\begin{minipage}[t][][c]{0.15\textwidth}
\includegraphics[width=7em]{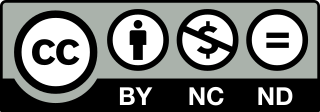}
\end{minipage}

\bibliography{references}





\appendix
\section{Estimating number of simulations with coefficient of variation}\label{appendix_CV}





\begin{figure*}[!htb]
    \centering
    \includegraphics[width=0.8\linewidth]{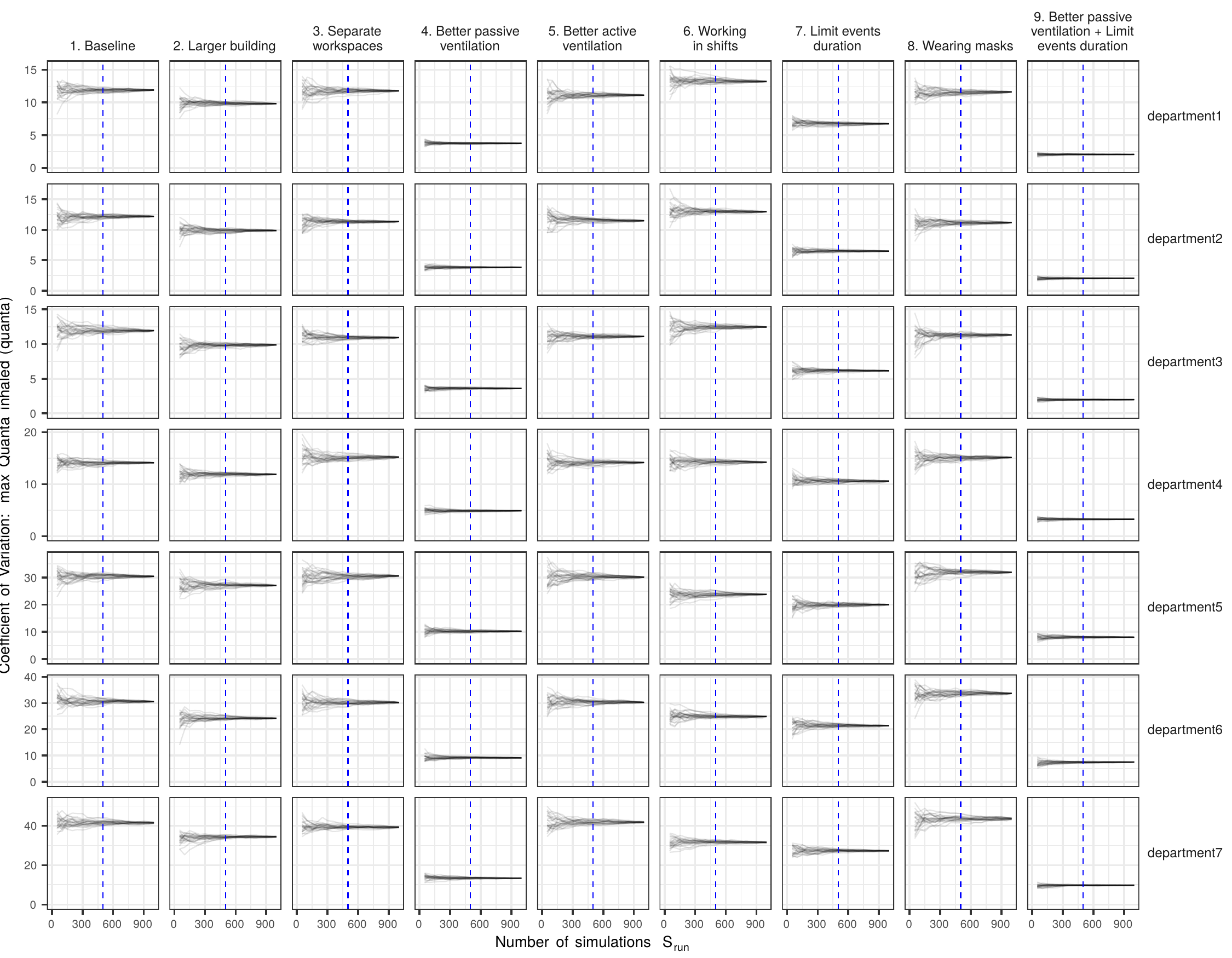} \\
    \includegraphics[width=0.8\linewidth]{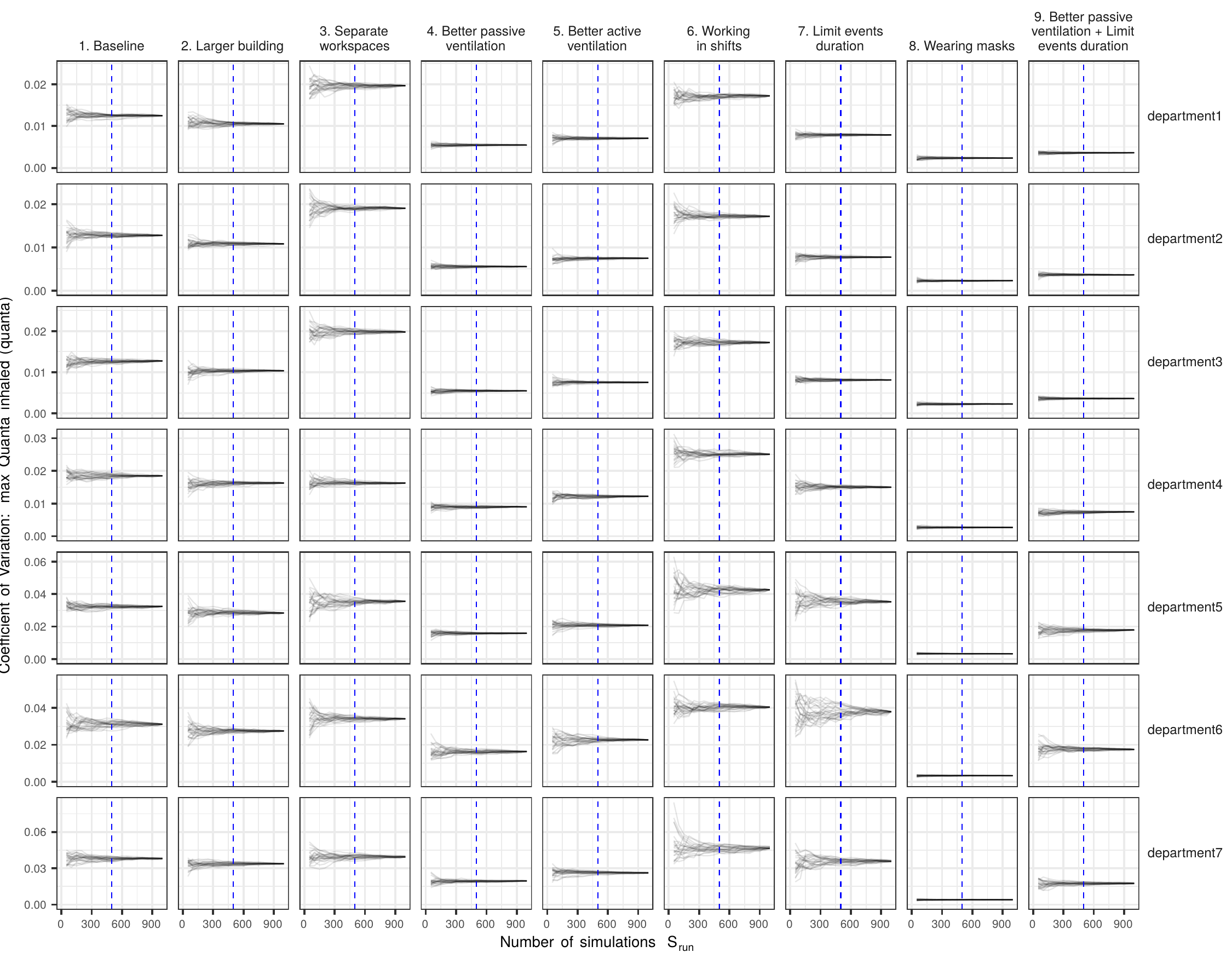}
    \caption{Results for determining an adequate number of simulation runs $S_{run}$. The \textit{Coefficient of Variation} (CV) was plotted for 100 repetitions of the same simulation configuration using different number of simulation runs. A stable CV was reached for $S_{run}=500$ simulation runs.}
    \label{fig:cv_people_quanta}
\end{figure*}

\end{document}